\documentclass[onecolumn]{aastex61}

\usepackage{graphicx}                  
\usepackage{color}                       

\usepackage[T1]{fontenc}
\usepackage[english]{babel}
\usepackage[utf8]{inputenc}
\usepackage{savesym}
\savesymbol{iint}
\savesymbol{iiint}
\usepackage{amsmath}
\usepackage{amsfonts}
\usepackage{amssymb}
\usepackage{tikz}
\usetikzlibrary{fadings}
\usepackage{enumitem}
\usepackage{float}
\allowdisplaybreaks

\chardef\us=`\_

\begin{document}

\title{Magneto-acoustic waves in a magnetic slab embedded in an asymmetric magnetic environment: Thin and wide slabs, hot and cold plasmas}

\correspondingauthor{R{\'o}bert Erd{\'e}lyi}
\email{robertus@sheffield.ac.uk}

\author{No{\'e}mi Kinga Zs{\'a}mberger}
\affiliation{Solar Physics and Space Plasma Research Centre, School of Mathematics and Statistics, University of Sheffield, Hicks Building, Hounsfield Road, Sheffield, S3 7RH, United Kingdom}
\affiliation{Department of Physics, University of Debrecen, 1. Egyetem t{\'e}r, H-4010, Debrecen, Hungary}
.

\author{R{\'o}bert Erd{\'e}lyi}
\affiliation{Solar Physics and Space Plasma Research Centre, School of Mathematics and Statistics, University of Sheffield, Hicks Building, Hounsfield Road, Sheffield, S3 7RH, United Kingdom}
\affiliation{Department of Astronomy, E{\"o}tv{\"o}s Lor{\'a}nd University, 1/A P{\'a}zm{\'a}ny P{\'e}ter s{\'e}t{\'a}ny, H-1117 Budapest, Hungary}

\begin{abstract}
Wave propagation in magnetically structured atmospheres is a tho\-rough\-ly studied, yet practically inexhaustible well of investigations in the field of solar magneto-seismology. A simple but powerful example is the examination of wave behaviour in a magnetic slab. Our previous study (Zs{\'a}mberger, Allcock and Erd{\'e}lyi, Astrophys. J., 853, p. 136, 2018) used an analytical approach to derive the general dispersion relation for magneto-acoustic waves in a magnetic slab of homogeneous plasma, which was enclosed in an asymmetric magnetic environment. In the present study, we focus on the analysis of wave propagation in various limiting cases applicable to solar and space plasma or astrophysics. The thin- and wide-slab approximations, as well as the limits of low and high plasma-$\beta$ values are considered. Utilising the fact that in a weakly asymmetric slab, the dispersion relation can be decoupled, the behaviour of quasi-sausage and quasi-kink modes is studied in further analytical and numerical detail, and their avoided crossings are described. The results highlight how the asymmetry influences the wave properties, e.g. the phase speed of eigenmodes, depending on the ratios of external-to-internal densities and magnetic fields on the two sides. Notably, the phase speeds of surface modes will converge to different values for quasi-sausage and quasi-kink modes in the wide-slab limit, and cut-off frequencies are introduced with respect to both surface and body modes, in thin as well as wide slabs, beyond which the solutions become leaky. These obtained properties of MHD wave behaviour could be measured with suitable high-resolution instruments in the future.
\end{abstract}


\section{Introduction} \label{sec:intro}

\par The existence of magnetohydrodynamic (MHD) waves in the solar atmosphere was predicted long before their actual detection \citep{uch-68, hab-79}. Contrary to the historically theoretical character of solar MHD research, nowadays, a wide variety of both space-based and ground-based instruments capable of unprecedented spatial and temporal resolution is available. Both Alfv{\'en} (see e.g. \citealt{jess-09}), fast \citep{mor-12}, and slow MHD waves \citep{fre-16} have been detected in the various features (e.g. coronal loops, prominences, sunspots) of the solar atmosphere and interpreted as oscillations in cylindrical (e.g. \citealt{asch-99}) or slab-like magnetised plasma configurations \citep{all-19}. This, in turn, motivates further analytical and numerical modelling in order to gain a better understanding of solar phenomena. This is the aim of solar magneto-seismology (SMS), which extends the scope of examinations by means of MHD waves from the corona (coronal seismology) to the lower parts of the magnetically coupled solar atmosphere \citep{somaseis1, somaseis2, obstrends}. 
\par Specifically, the study of slab geometry has a long history in SMS. A comprehensive discussion of the topic in a form popular today was given in three seminal articles of \citeauthor{roberts1} (\citeyear{roberts1, roberts2}) and \citeauthor{roberts3} (\citeyear{roberts3}). They revealed the details of linear wave propagation in a non-gravitational, (in)compressible, inviscid and ideal plasma. Their analysis found that the presence of a single interface may, under appropriate conditions, give rise to both the slow and the fast magnetoacoustic surface modes \citep{roberts1}. By introducing another interface, they constructed the model of a magnetic slab, which they \-examined first in field-free \citep{roberts2}, and then in a magnetic environment \citep{roberts3}. Some key steps and results in constructing and developing these slab models are summarised in \citet{all-18, asymag} and \citet{all-19}.
\par The, one may now label it as, classical model described by \citeauthor{roberts3} was symmetric about the centre of the slab. However, the solar atmosphere is a highly inhomogeneous medium with plenty of structuring, in which one cannot expect perfect symmetry to be present in the environment of MHD waveguides. Therefore, it was an important step forward in theoretical modelling when, as a generalisation of classical models, \citeauthor{asymm} (\citeyear{asymm}) introduced asymmetry into the slab geometry, by examining a magnetic slab is embedded in a non-magnetic but asymmetric environment. A further generalisation of the model was reached by dividing up the internal region into an arbitrary $N$ number of homogeneous slabs, as detailed by \citeauthor{shu-18} (\citeyear{shu-18}) and \citeauthor{all-19} (\citeyear{all-19}).  In our previous paper \citep{asymag}, we have explored the complexity and applicability of the slab model to a greater extent, by further generalising the slab model in a different manner, through embedding it in a magnetically asymmetric environment. We derived the general dispersion relation for linear perturbations and explored the fundamental effects of asymmetry on the nature of eigenmodes. We also carried out an application to magnetic bright points in the incompressible limit in order to demonstrate how powerful the analytical insight may be.
\par In the current paper, after a brief summary of the general results obtained in \citeauthor{asymag} (\citeyear{asymag}) necessary for the present work, we turn our attention to limiting cases that may be applicable to a number of solar and plasma-astrophysical structures. We suggest a few examples of such features that can be considered for magneto-seismological studies using the asymmetric slab model, however, the applicability of the model has to be evaluated on a case-by-case basis. First, the approximation of the equilibrium as a thin, and then as a wide slab are explored. Afterwards, the effect of the relationship between plasma parameters and the magnetic field is considered by examining the limits of zero (i.e. cold plasma), low, high, and infinite plasma-$\beta$ values. Finally, we explore the interesting phenomenon of avoided crossings shown by quasi-sausage and quasi-kink surface modes in response to varying key equilibrium parameters, such as e.g. density or magnetic field strength ratios between the slab and its environment.

\section{MHD waves in an asymmetric magnetic environment}  \label{sec:general}

\par
\begin{figure}[H]
\centering
\resizebox{0.75\linewidth}{!}{%
\begin{tikzpicture}
\path [fill=white!30!yellow!50!orange, opacity=0.65] (3.5,0) -- (3.5,4) -- (6.5,4) -- (6.5,0) -- (3.5,0);
\shade[left color=white!30!yellow!50!orange,right color=white!30!yellow!40!orange!25, opacity=0.5] (6.5,0) -- (6.5,4) -- (8.25,5) -- (8.25,0.85) -- (6.5,0);			
\shade[top color=white!30!yellow!50!orange,bottom color=white!30!yellow!40!orange!25, opacity=0.65] (3.5,0) -- (6.5,0) -- (6.5,-0.3) -- (3.5,-0.3) -- (3.5,0);		
\shade[top color=white!30!yellow!40!orange!25,bottom color=white!30!yellow!40!orange, opacity=0.65] (3.5,4) -- (5.25,5) -- (8.25,5) -- (6.5,4) -- (3.5,4);			
\shade[left color=white!30!yellow!50!orange,right color=white!30!yellow!40!orange!25, opacity=0.5] (6.5,0) -- (6.5,-0.3) -- (8.25,0.55) -- (8.25,0.85) -- (6.5,0.);		           

\path [fill=white!30!yellow!50!orange, opacity=0.25] (6.5,0) -- (6.5,4) -- (11,4) -- (11,0) -- (6.5,0);
\shade[top color=white!30!yellow!50!orange,bottom color=white!30!yellow!40!orange!25, opacity=0.25] (6.5,0) -- (11,0) -- (11,-0.3) -- (6.5,-0.3) -- (6.5,0);
\shade[top color=white!30!yellow!40!orange!25,bottom color=white!30!yellow!50!orange, opacity=0.25] (6.5,4) -- (8.25,5) -- (12.75,5) -- (11,4) -- (6.5,4);
\shade[left color=white!30!yellow!50!orange,right color=white!30!yellow!40!orange!25, opacity=0.25] (11,-0.3) -- (11,4) -- (12.75,5) -- (12.75,0.65) -- (11,-0.3);

\path [fill=white!30!yellow!50!orange, opacity=0.9] (-1,0) -- (-1,4) -- (3.5,4) -- (3.5,0) -- (-1,0);
\shade[top color=white!30!yellow!50!orange,bottom color=white!30!yellow!40!orange!25, opacity=0.9] (-1,0) -- (3.5,0) -- (3.5,-0.3) -- (-1,-0.3) -- (-1,0);
\shade[top color=white!30!yellow!40!orange!25,bottom color=white!30!yellow!50!orange, opacity=0.9] (-1,4) -- (0.75,5) -- (5.25,5) -- (3.5,4) -- (-1,4);

\draw [<->] (-1,1) -- (-1,0) -- (11,0);
\draw [->] (-1,0) -- (-0.5,0.3);

\draw [color=darkgray, ultra thick, dashed] (3.5,0) -- (3.5,4);
\draw [color=darkgray, ultra thick, dashed, path fading=east] (3.5,4) -- (5,4.9);
\draw [color=darkgray, ultra thick, dashed, path fading=east] (3.5,3) -- (4.5,3.6);
\draw [color=darkgray, ultra thick, dashed, path fading=east] (3.5,2) -- (4.5,2.6);
\draw [color=darkgray, ultra thick, dashed, path fading=east] (3.5,1) -- (4.5,1.6);
\draw [color=darkgray, ultra thick, dashed, path fading=east] (3.5,0) -- (5,0.9);


\draw [ultra thick, blue, path fading=north] (4.5,0) -- (4.5, 1.7); 
\draw [ultra thick, blue, path fading=south] (4.5,-0.3) -- (4.5,0);
\draw [ultra thick, blue, path fading=south] (4.5,2.2) -- (4.5,3.9);
\draw [ultra thick, blue, -stealth] (4.5,3.9) -- (4.5,4);

\draw [ultra thick, white!30!blue, path fading=north] (5,0.2) -- (5, 1.7); 
\draw [ultra thick, white!30!blue, path fading=south] (5,-0.1) -- (5,0.3);
\draw [ultra thick, white!30!blue, path fading=south] (5,2.2) -- (5,4.2);
\draw [ultra thick, white!30!blue, -stealth] (5,4.1) -- (5,4.3);

\draw [ultra thick, white!60!blue, path fading=north] (5.5,0.4) -- (5.5, 1.7); 
\draw [ultra thick, white!60!blue, path fading=south] (5.5,0.1) -- (5.5,0.6);
\draw [ultra thick, white!60!blue, path fading=south] (5.5,2.2) -- (5.5,4.5);
\draw [ultra thick, white!60!blue, -stealth] (5.5,4.3) -- (5.5,4.6);

\draw [ultra thick, white!75!blue, path fading=north] (6,0.6) -- (6, 1.7); 
\draw [ultra thick, white!75!blue, path fading=south] (6,0.3) -- (6,0.9);
\draw [ultra thick, white!75!blue, path fading=south] (6,2.2) -- (6,4.8);
\draw [ultra thick, white!75!blue, -stealth] (6,4.5) -- (6,4.9);

\draw [ultra thick, blue, path fading=north] (5.75,0) -- (5.75, 1.7);
\draw [ultra thick, blue, path fading=south] (5.75,-0.3) -- (5.75, 0);
\draw [ultra thick, blue, path fading=south] (5.75,2.2) -- (5.75,3.9);
\draw [ultra thick, blue, -stealth] (5.75,3.9) -- (5.75,4);

\draw [ultra thick, white!30!blue, path fading=north] (6.25,0.2) -- (6.25, 1.7);
\draw [ultra thick, white!30!blue, path fading=south] (6.25,-0.1) -- (6.25, 0.3);
\draw [ultra thick, white!30!blue, path fading=south] (6.25,2.2) -- (6.25,4.2);
\draw [ultra thick, white!30!blue, -stealth] (6.25,4.1) -- (6.25,4.3);

\draw [ultra thick, white!60!blue] (6.75,0.4) -- (6.75, 2.3);
\draw [ultra thick, white!60!blue, path fading=south] (6.75,0.1) -- (6.75, 0.6);
\draw [ultra thick, white!60!blue] (6.75,2.3) -- (6.75,4.5);
\draw [ultra thick, white!60!blue, -stealth] (6.75,4.3) -- (6.75,4.6);

\draw [ultra thick, white!75!blue] (7.25,0.6) -- (7.25, 2.6);
\draw [ultra thick, white!75!blue, path fading=south] (7.25,0.3) -- (7.25, 0.9);
\draw [ultra thick, white!75!blue] (7.25,2.6) -- (7.25,4.8);
\draw [ultra thick, white!75!blue, -stealth] (7.25,4.5) -- (7.25,4.9);

\draw [ultra thick, blue] (3.,0) -- (3., 2.2);
\draw [ultra thick, blue] (3.,2.2) -- (3.,3.9);
\draw [ultra thick, blue, path fading=south] (3.,-0.3) -- (3.,0);
\draw [ultra thick, blue, -stealth] (3.,3.9) -- (3.,4);

\draw [ultra thick, white!30!blue] (3.5,0.2) -- (3.5, 2.5);
\draw [ultra thick, white!30!blue] (3.5,2.5) -- (3.5,4.2);
\draw [ultra thick, white!30!blue, path fading=south] (3.5,-0.1) -- (3.5,0.3);
\draw [ultra thick, white!30!blue, -stealth] (3.5,4.1) -- (3.5,4.3);

\draw [ultra thick, white!60!blue, path fading=north] (4.,0.4) -- (4., 1.7);
\draw [ultra thick, white!60!blue, path fading=south] (4.,2.2) -- (4.,4.5);
\draw [ultra thick, white!60!blue, path fading=south] (4.,0.1) -- (4.,0.6);
\draw [ultra thick, white!60!blue, -stealth] (4.,4.3) -- (4.,4.6);

\draw [ultra thick, white!75!blue, path fading=north] (4.5,0.6) -- (4.5, 1.7);
\draw [ultra thick, white!75!blue, path fading=south] (4.5,2.2) -- (4.5,4.8);
\draw [ultra thick, white!75!blue, path fading=south] (4.5,0.3) -- (4.5,0.9);
\draw [ultra thick, white!75!blue, -stealth] (4.5,4.5) -- (4.5,4.9);

\draw [ultra thick, blue, path fading=north] (1.2,0) -- (1.2, 2.2);
\draw [ultra thick, blue, path fading=south] (1.2,2.8) -- (1.2,3.9);
\draw [ultra thick, blue, path fading=south] (1.2,-0.3) -- (1.2,0);
\draw [ultra thick, blue, -stealth] (1.2,3.9) -- (1.2,4);

\draw [ultra thick, white!30!blue, path fading=north] (1.7,0.2) -- (1.7, 2.2);
\draw [ultra thick, white!30!blue, path fading=south] (1.7,2.8) -- (1.7,4.2);
\draw [ultra thick, white!30!blue, path fading=south] (1.7,-0.1) -- (1.7,0.3);
\draw [ultra thick, white!30!blue, -stealth] (1.7,4.1) -- (1.7,4.3);

\draw [ultra thick, white!60!blue, path fading=north] (2.2,0.4) -- (2.2, 2.2);
\draw [ultra thick, white!60!blue, path fading=south] (2.2,2.8) -- (2.2,4.5);
\draw [ultra thick, white!60!blue, path fading=south] (2.2,0.1) -- (2.2,0.6);
\draw [ultra thick, white!60!blue, -stealth] (2.2,4.3) -- (2.2,4.6);

\draw [ultra thick, white!75!blue, path fading=north] (2.7,0.6) -- (2.7, 2.2);
\draw [ultra thick, white!75!blue, path fading=south] (2.7,2.8) -- (2.7,4.8);
\draw [ultra thick, white!75!blue, path fading=south] (2.7,0.3) -- (2.7,0.9);
\draw [ultra thick, white!75!blue, -stealth] (2.7,4.5) -- (2.7,4.9);

\draw [ultra thick, blue] (-0.6,0) -- (-0.6, 2.2);
\draw [ultra thick, blue] (-0.6,2.2) -- (-0.6,3.9);
\draw [ultra thick, blue, path fading=south] (-0.6,-0.3) -- (-0.6,0);
\draw [ultra thick, blue, -stealth] (-0.6,3.9) -- (-0.6,4);

\draw [ultra thick, white!30!blue] (-0.1,0.2) -- (-0.1, 2.5);
\draw [ultra thick, white!30!blue] (-0.1,2.5) -- (-0.1,4.2);
\draw [ultra thick, white!30!blue, path fading=south] (-0.1,-0.1) -- (-0.1,0.3);
\draw [ultra thick, white!30!blue, -stealth] (-0.1,4.1) -- (-0.1,4.3);

\draw [ultra thick, white!60!blue, path fading=north] (0.4,0.4) -- (0.4, 2.2);
\draw [ultra thick, white!60!blue, path fading=south] (0.4,2.8) -- (0.4,4.5);
\draw [ultra thick, white!60!blue, path fading=south] (0.4,0.1) -- (0.4,0.6);
\draw [ultra thick, white!60!blue, -stealth] (0.4,4.3) -- (0.4,4.6);

\draw [ultra thick, white!75!blue, path fading=north] (0.9,0.6) -- (0.9, 2.2);
\draw [ultra thick, white!75!blue, path fading=south] (0.9,2.8) -- (0.9,4.8);
\draw [ultra thick, white!75!blue, path fading=south] (0.9,0.3) -- (0.9,0.9);
\draw [ultra thick, white!75!blue, -stealth] (0.9,4.5) -- (0.9,4.9);


\draw [ultra thick, blue, path fading=south] (7.25,-0.3) -- (7.25,0.2);
\draw [ultra thick, blue] (7.25,0.2) -- (7.25,3.9);
\draw [ultra thick, blue, -stealth] (7.25,3.9) -- (7.25,4);

\draw [white!30!blue, ultra thick, path fading=north] (7.75,0.2) -- (7.75, 2.2);
\draw [white!30!blue, ultra thick, path fading=south] (7.75,2.8) -- (7.75,4.2);
\draw [white!30!blue, ultra thick, path fading=south] (7.75,-0.1) -- (7.75,0.3);
\draw [white!30!blue, ultra thick, -stealth] (7.75,4.1) -- (7.75,4.3);

\draw [white!60!blue, ultra thick, path fading=north] (8.25,0.4) -- (8.25, 2.2);
\draw [white!60!blue, ultra thick, path fading=south] (8.25,2.8) -- (8.25,4.5);
\draw [white!60!blue, ultra thick, path fading=south] (8.25,0.1) -- (8.25,0.6);
\draw [white!60!blue, ultra thick, -stealth] (8.25,4.3) -- (8.25,4.6);

\draw [white!75!blue, ultra thick, path fading=north] (8.75,0.6) -- (8.75, 2.2);
\draw [white!75!blue, ultra thick, path fading=south] (8.75,2.8) -- (8.75,4.8);
\draw [white!75!blue, ultra thick, path fading=south] (8.75,0.3) -- (8.75,0.9);
\draw [white!75!blue, ultra thick, -stealth] (8.75,4.5) -- (8.75,4.9);

\draw [ultra thick, blue, path fading=north] (8.5,0) -- (8.5, 2.2);
\draw [ultra thick, blue, path fading=south] (8.5,2.8) -- (8.5,3.9);
\draw [ultra thick, blue, path fading=south] (8.5,-0.3) -- (8.5,0);
\draw [ultra thick, blue, -stealth] (8.5,3.9) -- (8.5,4);

\draw [white!30!blue, ultra thick, path fading=north] (9,0.2) -- (9, 2.2);
\draw [white!30!blue, ultra thick, path fading=south] (9,2.8) -- (9,4.2);
\draw [white!30!blue, ultra thick, path fading=south] (9,-0.1) -- (9,0.3);
\draw [white!30!blue, ultra thick, -stealth] (9,4.1) -- (9,4.3);

\draw [white!60!blue, ultra thick, path fading=north] (9.5,0.4) -- (9.5, 2.2);
\draw [white!60!blue, ultra thick, path fading=south] (9.5,2.8) -- (9.5,4.5);
\draw [white!60!blue, ultra thick, path fading=south] (9.5,0.1) -- (9.5,0.6);
\draw [white!60!blue, ultra thick, -stealth] (9.5,4.3) -- (9.5,4.6);

\draw [white!75!blue, ultra thick, path fading=north] (10,0.6) -- (10, 2.2);
\draw [white!75!blue, ultra thick, path fading=south] (10,2.8) -- (10,4.8);
\draw [white!75!blue, ultra thick, path fading=south] (10,0.3) -- (10,0.9);
\draw [white!75!blue, ultra thick, -stealth] (10,4.5) -- (10,4.9);

\draw [ultra thick, blue, path fading=north] (9.75,0) -- (9.75, 2.2);
\draw [ultra thick, blue, path fading=south] (9.75,2.8) -- (9.75,3.9);
\draw [ultra thick, blue, path fading=south] (9.75,-0.3) -- (9.75,0);
\draw [ultra thick, blue, -stealth] (9.75,3.9) -- (9.75,4);

\draw [white!30!blue, ultra thick, path fading=north] (10.25,0.2) -- (10.25, 2.2);
\draw [white!30!blue, ultra thick, path fading=south] (10.25,2.8) -- (10.25,4.2);
\draw [white!30!blue, ultra thick, path fading=south] (10.25,-0.1) -- (10.25,0.3);
\draw [white!30!blue, ultra thick, -stealth] (10.25,4.1) -- (10.25,4.3);

\draw [white!60!blue, ultra thick, path fading=south] (10.75,0.1) -- (10.75,0.6);
\draw [white!60!blue, ultra thick] (10.75,0.6) -- (10.75,4.5);
\draw [white!60!blue, ultra thick, -stealth] (10.75,4.3) -- (10.75,4.6);

\draw [white!75!blue, ultra thick, path fading=south] (11.25,0.3) -- (11.25,0.8);
\draw [white!75!blue, ultra thick] (11.25,0.8) -- (11.25,4.8);
\draw [white!75!blue, ultra thick, -stealth] (11.25,4.5) -- (11.25,4.9);

\draw [ultra thick, blue] (11.,0.2) -- (11.,3.9);
\draw [ultra thick, blue, path fading=south] (11.,-0.3) -- (11.,0.2);
\draw [ultra thick, blue, -stealth] (11.,3.9) -- (11.,4);

\draw [white!40!blue, ultra thick] (11.5,0.4) -- (11.5,4.2);
\draw [white!40!blue, ultra thick, path fading=south] (11.5,-0.1) -- (11.5,0.4);
\draw [white!40!blue, ultra thick, -stealth] (11.5,4.1) -- (11.5,4.3);

\draw [white!70!blue, ultra thick] (12.,0.4) -- (12.,4.5);
\draw [white!70!blue, ultra thick, path fading=south] (12.,0.1) -- (12.,0.6);
\draw [white!70!blue, ultra thick, -stealth] (12.,4.3) -- (12.,4.6);

\draw [white!80!blue, ultra thick] (12.5,0.9) -- (12.5,4.8);
\draw [white!80!blue, ultra thick, path fading=south] (12.5,0.3) -- (12.5,0.9);
\draw [white!80!blue, ultra thick, -stealth] (12.5,4.5) -- (12.5,4.9);

\draw [color=darkgray, ultra thick, dashed] (6.5,0) --(6.5,4);
\draw [color=darkgray, ultra thick, dashed, path fading=east] (6.5,4) -- (8,4.9);
\draw [color=darkgray, ultra thick, dashed, path fading=east] (6.5,2) -- (7.5,2.6);
\draw [color=darkgray, ultra thick, dashed, path fading=east] (6.5,1) -- (7.5,1.6);
\draw [color=darkgray, ultra thick, dashed, path fading=east] (6.5,3) -- (7.5,3.6);
\draw [color=darkgray, ultra thick, dashed, path fading=east] (6.5,0) -- (8,0.9);

\small
\node [below] at (3.5,0) {$-x_0$};
\node [below] at (6.5,0) {$x_0$};
\node [below] at (10.75,0) {$x$};
\node [left] at (-0.55,1) {$z$};
\node [right] at (-0.5,0.3) {$y$};

\large
\node [right] at (0.15,2.5) {$\rho_1$, $p_1$, $T_1$, $B_1$};
\node [right] at (3.75,2) {$\rho_0$, $p_0$, $T_0$, $B_0$};
\node [right] at (7.6,2.5) {$\rho_2$, $p_2$, $T_2$, $B_2$};
\end{tikzpicture}
}
\caption{The equilibrium: a magnetic slab, $|x|\leq{}x_0$ (medium orange colour), sandwiched between two, semi-infinite uniform magnetised plasmas, $x<-x_0$ and $x>x_0$ (light and dark orange). The blue arrows illustrate the magnetic fields, $B_0\mathbf{\hat{z}}$,  $B_1\mathbf{\hat{z}}$ and  $B_2\mathbf{\hat{z}}$; and the dashed black lines outline the boundaries of the slab.}
\label{fig:eq}
\end{figure}
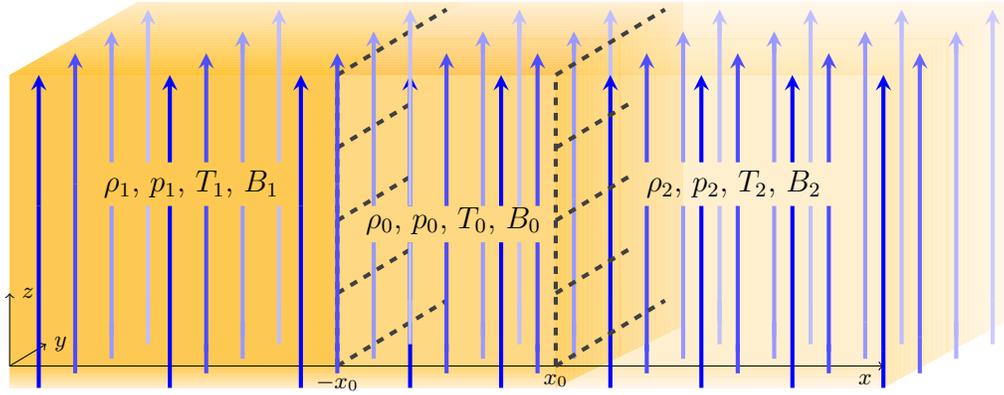

 We investigate the magnetic waveguide model comprised of unbounded, three-di\-men\-sio\-nal, inviscid and ideal plasma embedded in equilibrium magnetic field $B_0(x)\mathbf{\hat{z}}$, where $\mathbf{\hat{z}}$ is the unit vector in the vertical direction. In order to make the influence of the magnetic asymmetry itself clear, only magneto-acoustic waves are studied, and therefore, the effects of gravity and background bulk motions are not considered. The volume is divided by two surfaces of discontinuity, defining three domains of uniform plasma, with different densities, $\rho$, pressures, $p$, temperatures, $T$, and magnetic field strengths, $B$, across the domains:
\begin{equation}
  N(x)=\begin{cases}
    N_1 & \qquad x<-x_0,\\
    N_0 & \qquad |x|<x_0, \\
    N_2 & \qquad x_0< x,\\
  \end{cases} 
\end{equation}			
where $N_i$ denotes any of the physical parameters listed above, namely $N_i = \text{ constant }$ (for $i=0,1,2$). An illustration of this equilibrium configuration can be found in Figure \ref{fig:eq}.
\par Disturbances in the slab and its environment are governed by the ideal MHD equations. By performing a linearisation, and constraining the study to plane-wave solutions propagating in the $z$-direction (i.e. along the slab), we determined that each domain ($i=0,1,2$) is governed by an ordinary differential equation of the form
\begin{equation}
	\hat{v}_{x}''-m_i^2 \hat{v}_{x} = 0,		\label{theode}
\end{equation}
where $\hat{v}_{x}$ is the amplitude of the $x$-component of the velocity perturbation introduced, and
\begin{equation}
	m_i^2=\frac{\left( k^2 v_{Ai}^2 - \omega^2\right)\left( k^2 c_i^2 - \omega^2\right)}{\left(v_{Ai}^2 + c_i^2\right)\left( k^2 c_{Ti}^2 - \omega^2\right)}.
\end{equation}

Here, $\omega$ is the angular frequency of the waves, and $k$ is the $z$-component of the wavenumber vector.  The characteristic speeds in the plasma are: the Alfv{\'e}n speed, $v_{Ai}=B_i/\sqrt{\rho_i \mu}$, where $\mu$ is the permeability of free space; and the sound speed, $c_i=\sqrt{\gamma p_i/\rho_i}$, where $\gamma$ is the adiabatic index. $\mu$ and $\gamma$ are uniform across all the domains, as the plasma composition is assumed to be the same in the entire configuration. The third characteristic speed,
\begin{equation}
c_{Ti}^2 = \frac{v_{Ai}^2  c_i^2}{v_{Ai}^2 + c_i^2}, 
\end{equation}
is the so-called cusp or tube speed of a given domain, which is a sub-sonic and sub-Alfv{\'e}nic speed. 
\par For physically real solutions that are evanescent outside the slab, following \citeauthor{asymag} (\citeyear{asymag}), we found the dispersion relation to be
\begin{align}
& 2 \frac{\rho_0}{\rho_1} m_1 \frac{\rho_0}{\rho_2} m_2 \left( k^2 v_{A0}^2 - \omega^2\right)^2 + 2 m_0^2 \left( k^2 v_{A1}^2 - \omega^2\right) \left( k^2 v_{A2}^2 - \omega^2\right) \nonumber\\ 
& \quad + \rho_0 m_0 \left( k^2 v_{A0}^2 - \omega^2\right) \left[ \frac{m_2}{\rho_2} \left( k^2 v_{A1}^2 - \omega^2\right) + \frac{m_1}{\rho_1} \left( k^2 v_{A2}^2 - \omega^2\right) \right]  \left[ \tau_0 + \frac{1}{\tau_0} \right] =0, \label{fullsurface}
\end{align}
where $\tau_0 = \tanh{(m_0 x_0)}$. It is apparent that the full dispersion relation does not decouple into separate solutions for sausage or kink modes, as it would in the symmetric case. Accordingly, the eigenmodes show mixed properties, which is why we refer to them as quasi-sausage and quasi-kink modes (see also \citeauthor{asymm} \citeyear{asymm}). If the asymmetry is weak, i.e. the pressures, densities and magnetic field strengths do not differ too strongly on the two sides of the slab, the dispersion relation decouples into two equations:
\begin{align}
 (k^2 v_{A0}^2-\omega^2)  \left[ \frac{ \rho_0}{\rho_1} \frac{m_1}{  (k^2 v_{A1}^2-\omega^2)} + \frac{ \rho_0}{\rho_2}  \frac{m_2}{  (k^2 v_{A2}^2-\omega^2)} \right] + 2   m_0  \binom{\tanh}{\coth} \{m_0 x_0\} = 0, \label{surface}
\end{align}
where the substitution of $\tanh{(m_0 x_0)}$ describes quasi-sausage modes, and $\coth{(m_0 x_0)}$ gives quasi-kink mode solutions. In the following sections, these dispersion relations will be further examined in limits that are often used in solar or plasma-astrophysics.

\section{Thin-slab approximation} 	\label{sec:thin} 

In the thin-slab approximation, the wavelength, $\lambda$, of the waves is much greater than the width of the slab: $x_0/\lambda \approx kx_0 \ll 1$. This limit may have both photospheric or coronal applications, if we describe them in Cartesian rather than cylindrical geometry. Such a description may be applicable to various solar phenomena, such as prominences (see \citeauthor{arr-oli-bal-12} \citeyear{arr-oli-bal-12}), sunspot light bridges and light walls \citep{yua-nak-14, yan-zha-16, yan-zha-17}, magnetic bright points \citep{utz-09, liu-18}, or any thin and magnetised plasma-astrophysical object that is sandwiched between uniform, homogeneous but asymmetric magnetised semi-infinite plasma environments as a first approximation.

\subsection{Surface modes} 	\label{sec:ThinSurface} 

\par We have only considered perturbations that are evanescent outside the slab, but it should be noted that surface modes are evanescent inside the slab as well, mostly perturbing regions close to the slab boundaries. 

\subsubsection{Quasi-sausage surface modes} 	\label{sec:QS-ThinSurface} 

First, let us examine quasi-sausage surface modes, which are described by the component of Equation \eqref{surface} containing the odd $\tanh{(m_0 x_0)}$ function. Supposing that, in this limit, $m_0 x_0 \ll 1$, it follows that $\tanh{m_0 x_0} \approx m_0 x_0$. Substituting this into equation \eqref{surface}, the dispersion relation for quasi-sausage surface modes becomes
\begin{align}
 (k^2 v_{A0}^2-\omega^2)  \left[ \frac{ \rho_0}{\rho_1} \frac{m_1} { (k^2 v_{A1}^2-\omega^2) } +  \frac{ \rho_0}{\rho_2}  \frac{m_2}{  (k^2 v_{A2}^2-\omega^2)} \right] + 2  m_0^2 x_0 = 0. \label{surfacethin}
\end{align}
The frequency $\omega^2 = k^2 v_{A0}^2$ would be a trivial solution not considered here (for reasons see \citeauthor{asymag} \citeyear{asymag}). One group of solutions might occur when the phase speed of the waves approaches the cusp speed: $\omega^2 \rightarrow k^2 c_{T0}^2$. Substitution of this approximation into (\ref{surfacethin}), after some algebra, yields
\begin{align}
         \omega^2 = k^2 c_{T0}^2 \left[ 1 + \frac{2 (c_0^2 - c_{T0}^2) (v_{A1}^2-c_{T0}^2)^{1/2} (v_{A2}^2-c_{T0}^2)^{1/2}k x_0}{\rho_0 v_{A0}^2 c_0^2 R_v} \right], \nonumber 
\label{eq:ss1}
\end{align}
where
\begin{align}
       R_v& = \frac{1}{\rho_2} \frac{(v_{A1}^2-c_{T0}^2)^{1/2} (c_{2}^2-c_{T0}^2)^{1/2}}{(v_{A2}^2+c_{2}^2)^{1/2} (c_{T2}^2-c_{T0}^2)^{1/2}} + \frac{1}{\rho_1} \frac{(v_{A2}^2-c_{T0}^2)^{1/2} (c_{1}^2-c_{T0}^2)^{1/2}}{(v_{A1}^2+c_{1}^2)^{1/2} (c_{T1}^2-c_{T0}^2)^{1/2}}  .
\end{align}
\par This wave solution is a slow quasi-sausage surface mode, which nears $\omega^2 \rightarrow k^2 c_{T0}^2$ from above as $k x_0 \rightarrow 0$ (the slab becomes thinner). 
The condition for its existence is, without any further information on the values of characteristic speeds on either side of the slab, the following:
\begin{align}
&\sqrt{ c_{T1}^2- c_{T0}^2} > 0 \Rightarrow  c_{T0}^2 < c_{T1}^2  \qquad \text{and } \quad \sqrt{ c_{T2}^2- c_{T0}^2} > 0 \Rightarrow c_{T0}^2 < c_{T2}^2.	\label{ctconditions}
\end{align}
\par The effect on further possible characteristic speed orderings on this group of solutions is examined in Section \ref{sec:appendix-tube} of the Appendix.

\par A different type of quasi-sausage mode solutions approaches one of the external sound speeds in the thin-slab limit. For example, if we take the approximation $\omega^2 \rightarrow  k^2 c_2^2$, the solutions are given by
\begin{align}
\omega^2 &= k^2 c_2^2 - \left[ \frac{\rho_2}{\rho_0}\frac{ 2 (c_{T2}^2 - c_2^2)^{1/2} (v_{A2}^4 - c_2^4)^{1/2} (c_0^2 - c_2^2) k^2 x_0}{ (c_{T0}^2 - c_2^2) (c_0^2 + v_{A0}^2)}  + \frac{\rho_2}{\rho_1}\frac{ (c_{T2}^2 - c_2^2)^{1/2} (v_{A2}^4 - c_2^4)^{1/2} (c_{1}^2 - c_2^2)^{1/2} }{ (c_{T1}^2 - c_2^2)^{1/2} (v_{A1}^2 - c_2^2)^{1/2} (v_{A1}^2 + c_1^2)^{1/2} }  \right]^2.
\label{eq:ss2}
\end{align}
This surface wave solution exists when $c_2 < c_{T1}$ or $\min{(c_1, v_{A1})} < c_2 < \max{(c_1, v_{A1})}$, since outside these bounds, the waves would become leaky. Naturally, the same type of solution can be found if the indices $j=1,2$ are swapped.  
\par Let us now consider the case with an isothermal external environment, i.e. when the external sound speeds are the same:  $c_1^2=c_2^2=c_e^2$, the solutions are derived by substituting $\omega^2 \approx k^2 c_e^2$ into Equation \eqref{surfacethin}, yielding
\begin{align}
\omega^2 &= k^2 \left[ c_e^2 + \frac{4 (c_0^2 - c_e^2)^2 (k x_0) ^2 }{\rho_0^2 (v_{A0}^2 + c_0^2)^2 (c_{T0}^2 - c_e^2)^2 R_v^2} \right], \nonumber \\
R_v^2 &= \left[ \frac{1}{\rho_2} \frac{1}{(v_{A1}^2-c_e^2)^{1/2} (c_e^2+ v_{A2}^2)^{1/2} (c_{T2}^2-c_e^2)^{1/2}}  + \frac{1}{\rho_1} \frac{1}{(v_{A2}^2-c_e^2)^{1/2} (c_e^2+ v_{A1}^2)^{1/2} (c_{T1}^2-c_e^2)^{1/2}} \right]^2, \label{eq:ss3}
\end{align}
for $v_{A1}, v_{A2} < c_{e}$ and  $c_{e} < c_{T1}, c_{T2}$. Supposing that $v_{A1}^2 = v_{A2}^2 = v_{Ae}^2$, then $\rho_1 = \rho_2 = \rho_e$ has to be true as well, which leads back to Equation (16a) of \citeauthor{roberts3} (\citeyear{roberts3}). If the external plasma environment is non-magnetic, this case further reduces to Equation (32) of \citeauthor{asymm} (\citeyear{asymm}).

\subsubsection{Quasi-kink surface modes} 	\label{sec:QK-ThinSurface}

\par Let us now consider quasi-kink mode solutions, which are governed by the $\coth{(m_0 x_0)}$ part of the decoupled dispersion relation (Equation \ref{surface}). In the limit of $m_0 x_0 \ll 1$,  $\coth{m_0 x_0} \approx (m_0 x_0)^{-1}$. Substituting this into (\ref{surface}), the dispersion relation for quasi-kink modes becomes
\begin{equation}
 \rho_0 x_0 (k^2 v_{A0}^2-\omega^2)  \left[ \frac{ m_1}{\rho_1  (k^2 v_{A1}^2-\omega^2)} + \frac{ m_2}{\rho_2  (k^2 v_{A2}^2-\omega^2) }\right] + 2 = 0. \label{kinkthin}
\end{equation}
One kind of these modes might approach one of the external Alfv{\'e}n speeds in the thin-slab approximation. We can obtain this solution by substituting the limit $\omega^2 \rightarrow k^2 v_{A1}^2$ into Equation \eqref{kinkthin}:
\begin{align}
\omega^2 = k^2 \left[ v_{A1}^2 - \frac{\rho_0^2 \rho_2^2}{\rho_1^2} \frac{(c_1^2 - v_{A1}^2) (v_{A0}^2 - v_{A1}^2)^2 (v_{A2}^2 - v_{A1}^2) (c_{T2}^2 - v_{A1}^2) (k^2 x_0)^2 }{(c_{T1}^2 - v_{A1}^2) R_v^2} \right], \label{eq:sk1}
\end{align}
where, now, 
\begin{align}
R_v &=  2 \rho_2 k (v_{A2}^2 - v_{A1}^2)^{1/2} (c_{T2}^2 - v_{A1}^2 )^{1/2} (v_{A1}^2 + c_1^2)^{1/2}   + \rho_0 (v_{A0}^2 - v_{A1}^2) (c_2^2 - v_{A1}^2)^{1/2} k^2 x_0 . \nonumber
\end{align}
This mode exists as a trapped perturbation when $v_{A1}^2 < c_{T2}^2$ or $ \min{(v_{A2}^2, c_2^2)} < v_{A1}^2 < \max{(v_{A2}^2, c_2^2)}$. When $v_{A1}^2 = v_{A2}^2 = v_{Ae}^2$, the solution further simplifies to
\begin{align}
\omega^2 = k^2 v_{Ae}^2 \left[ 1- \left( 1- \frac{v_{A0}^2}{v_{Ae}^2}\right)^2 \left(\frac{\rho_0 (k x_0)}{2} \right)^2 \left( \frac{1}{\rho_2} \sqrt{1- \frac{c_2^2}{v_{Ae}^2}} + \frac{1}{\rho_1} \sqrt{1- \frac{c_1^2}{v_{Ae}^2}}\right)^2 \right].
\label{eq:sk2}
\end{align}
In the case of an isothermal external environment, i.e. $c_1^2=c_2^2=c_e^2$, and so $\rho_1=\rho_2=\rho_e$, the obtained solution leads back to the one for the symmetric slab (Equation (18a) of \citeauthor{roberts3} \citeyear{roberts3}). 
\par An asymmetric equivalent for a different type of kink-mode solutions can be found as well, namely, for those that approach one of the external cusp speeds. With the substitution $\omega^2 \rightarrow k^2 c_{T1}^2$, Equation \eqref{kinkthin} becomes
\begin{align}
\omega^2 &= k^2 \left[ c_{T1}^2 - \frac{\rho_0^2 \rho_2^2}{\rho_1^2}  \frac{ R_{v1} (k^2 x_0)^2}{(v_{A1}^2 - c_{T1}^2) (c_1^2 + v_{A1}^2) R_{v2}^2} \right], \label{eq:sk3}
\end{align}
with
\begin{align}
R_{v1}&= (c_1^2 - c_{T1}^2)(v_{A0}^2 - c_{T1}^2)^2 (v_{A2}^2 - c_{T1}^2) (c_{T2}^2 - c_{T1}^2) (v_{A2}^2 + c_2^2), \nonumber \\
R_{v2}&= 2 \rho_2 k (v_{A2}^2 - c_{T1}^2 )^{1/2} (c_{T2}^2 - c_{T1}^2)^{1/2} (v_{A2}^2 + c_{2}^2)^{1/2} + \rho_0 k^2 x_0 (v_{A0}^2 - c_{T1}^2) (c_2^2 - c_{T1}^2)^{1/2}. \nonumber
\end{align}
This solution is a trapped oscillation when $c_{T1}^2 < c_{T2}^2$ or  $ \min{(v_{A2}^2, c_2^2)} < c_{T1}^2 < \max{(v_{A2}^2, c_2^2)}$. When the two external cusp speeds are the same, this case reduces to  Equation (18b) of \citeauthor{roberts3} (\citeyear{roberts3}).
An asymmetrised generalisation of \citeauthor{roberts3}'s (\citeyear{roberts3}) Equation (19), the approximation for the case when $v_{Ae}/v_{A0}$ is of the order of $kx_0$ can also be obtained:
\begin{align}
\omega^2 = k^2 v_{A1}^2 \left[1 + \frac{\rho_0 \rho_2}{\rho_1} \frac{v_{A0}^2}{v_{A1}^2} \frac{v_{A2}^2 (kx_0)}{2 \rho_2 v_{A2}^2 + \rho_0 v_{A0}^2 x_0}  \right]
\end{align}
if $v_{A1} \ll v_{A2}$ is also satisfied. If, conversely, $v_{A2} \ll v_{A1}$, the solution becomes
\begin{align}
\omega^2 = k^2 v_{A1}^2 \left[1 + \frac{\rho_0 \rho_2}{\rho_1} \frac{v_{A0}^2}{v_{A1}^2} \frac{v_{A1}^2 (kx_0) }{\rho_0 v_{A0}^2 x_0 - 2 \rho_2 v_{A1}^2}  \right].
\end{align}
When $v_{A1}^2 = v_{A2}^2 = v_{Ae}^2$ holds, this approximation may be given as
\begin{align}
\omega^2 = k^2 v_{Ae}^2  \left[ 1 + \frac{1}{R} \frac{v_{A0}^2}{v_{Ae}^2 } (kx_0) \right],
\end{align}
where
\begin{align}
R= \left[ \frac{\rho_0}{2} \left( \frac{1}{\rho_1} + \frac{1}{\rho_1} \right) \right]^{-1}
\end{align}
is the measure of the density asymmetry used in \citeauthor{asymag} (\citeyear{asymag}). 
\par Equations \eqref{eq:ss1}-\eqref{eq:ss3} and \eqref{eq:sk1}-\eqref{eq:sk3} show us that the overall structure of the solutions in the thin-slab limit of an asymmetric magnetic slab remains similar to the symmetric case. This actually confirms how powerful the initial model of a symmetric slab is, which may be seen as practical tool when interpreting MHD wave observations. While analytical approximations of the solutions can still be given, wave dispersion in the asymmetric configuration, however, becomes more complex. The differences in environmental equilibrium parameters can introduce cut-off frequencies, beyond which the oscillations become leaky. In general, Equations \eqref{eq:ss1}-\eqref{eq:sk3} also reveal that surface waves in the magnetic slab are quite sensitive to the relative magnitudes of external densities compared to the internal one, which is why they can be shown to possess avoided crossings (see Section \ref{sec:avcross}).

\subsection{Body modes}		\label{sec:ThinBody} 

\par Still in the thin-slab approximation, let us now examine the existence and characteristics of body waves. First of all, the dispersion relation itself can be rewritten without the use of hyperbolic functions. As opposed to surface waves, where $m_0^2$  was positive, in the case of body waves, $m_0^2 < 0$. Defining $n_0^2 := - m_0^2 > 0$, the dispersion relation (Equation \ref{fullsurface}) becomes now:
\begin{align}
& 2 \frac{\rho_0}{\rho_1} m_1 \frac{\rho_0}{\rho_2} m_2 \left( k^2 v_{A0}^2 - \omega^2\right)^2 - 2 n_0^2 \left( k^2 v_{A1}^2 - \omega^2\right) \left( k^2 v_{A2}^2 - \omega^2\right) + \nonumber \\
& \rho_0 n_0 \left( k^2 v_{A0}^2 - \omega^2\right) \left[ \frac{m_1}{\rho_1} \left( k^2 v_{A2}^2 - \omega^2\right)   + \frac{m_2}{\rho_2 }\left( k^2 v_{A1}^2 - \omega^2\right)  \right]  \left[-\tan{n_0 x_0} +  \cot{n_0 x_0} \right] =0  \label{fullbody} .
\end{align}
\par Here, not only the full, but also the decoupled counterpart of the dispersion relation (Equation \ref{surface}) may be expressed with the tangent and cotangent functions as
\begin{equation}
 (k^2 v_{A0}^2-\omega^2)  \left[ \frac{ \rho_0}{\rho_1} \frac{m_1}{(k^2 v_{A1}^2-\omega^2)}+ \frac{ \rho_0}{\rho_2}  \frac{m_2}{(k^2 v_{A2}^2-\omega^2)}\right] + 2  n_0 \binom{-\tan}{\cot} \{n_0 x_0\} = 0. \label{body}
\end{equation}
\par  Finding body mode solutions generally requires different considerations than those used above for surface modes, since assuming that $m_0 x_0 \rightarrow 0 $ as the slab becomes thinner ($k x_0 \rightarrow 0$) will not describe every possible wave mode \citep{roberts2}. Let us prescribe therefore that $m_0 x_0$ should remain \textit{bounded} as $k x_0 \rightarrow 0$ tends towards zero. Considering the dispersion relation for quasi-sausage body waves, the expression $n_0 \tan{(n_0 x_0)}$  needs to remain finite. This necessitates that $n_0 x_0$ converge to the roots of $\tan{(n_0 x_0)} = 0$, that is, $n_0 x_0 = j \pi $ (for $j=1,2,3$ ...). Substituting $\omega^2 \approx k^2 c_{T0}^2 (1 + \nu (k x_0)^2)$ into the definition of $n_0$ and multiplying by $x_0$, we can find the values of $\nu$ as follows:
\begin{align}
n_0^2 x_0^2 &= - m_0^2 x_0^2 =  \frac{(c_0^2-c_{T0}^2) (v_{A0}^2 - c_{T0}^2) }{(c_0^2 + v_{A0}^2) c_{T0}^2  \nu }. \label{sbthinnu}
\end{align}
Due to the condition on the values of $n_0^2 x_0^2$, this also equals $j^2 \pi^2$. Substituting this expression and rearranging the equation yields $\nu$ for every (integer) $j$:
\begin{align}
\nu_j=  \frac{(c_0^2-c_{T0}^2) (v_{A0}^2 - c_{T0}^2) }{(c_0^2 + v_{A0}^2) c_{T0}^2   j^2 \pi^2}. \label{nutan}
\end{align}
\par We have thus found that there are countably many quasi-sausage body mode solutions, with a different number of nodes inside the slab, which we will call harmonics in the direction of structuring, or, in short, harmonics. The situation so far is analogue algebraically to that in the asymmetric slab in a field-free environment \citep{asymm}. This type of description so far does not deal with the influence that the difference in external equilibrium parameters has on the slab system. There are two possibilities to provide an approximation that considers the effects of external magnetic asymmetry. For example, it is conceivable that either of the external sound or Alfv{\'e}n speeds being higher than $c_{T0}$ may introduce a cut-off frequency, which prevents the phase speed from converging to the cusp speed in the limit of a thin slab.
\par In the dispersion relation for body modes (\ref{body}), the coefficients $n_0^2, m_1^2, m_2^2$ all must \textit{simultaneously} have positive values. In adherence with these requirements, there are three possibilities for slow body mode waves to exist:
\begin{subequations}
\begin{alignat}{1}
&\max{[c_{T0}, \min{(c_1, v_{A1})}, \min{(c_2, v_{A2})}]} < v_{ph}   <  \min{}[ \min{(c_0, v_{A0})},  \max{(c_1, v_{A1})},   \max{}(c_2, v_{A2}) ], \label{sba}\\
&\max{[c_{T0}, \min{(c_1, v_{A1})}]} < v_{ph} < \min{[ \min{(c_0, v_{A0})}, \max{(c_1, v_{A1})},c_{T2} ]}, \label{sbb}\\
 &c_{T0} < v_{ph} < \min{[ \min{(c_0, v_{A0})}, c_{T1},  c_{T2} ]}. \label{sbc}
\end{alignat}
\end{subequations}
An additional fourth category could be defined by swapping the $i=1,2$ indices in condition \eqref{sbb}. We will, however, not deal with this case in further detail, since it does not describe a qualitatively different type of body mode, and one need only swap the same indices in the description of the solution curves that belong to condition \eqref{sbb}, in order to obtain the solutions for such a mirrored situation. The same will be true for the phase speed bands allowing the existence of fast body modes in the thin-slab approximation, as well as the bands of both slow and fast body waves in the wide-slab limit.
\par Proceeding from here, one possibility is to use Equation \eqref{sbthinnu}, and only accept the solutions while they are in either one of the phase speed bands delineated in Equations \eqref{sba} - \eqref{sbc}. Another approach, which we will follow now, is to use an approximation which bounds the solutions to remain in the above-mentioned bands. One must, however, remember that in the extremes of the thin-slab limit, solutions can become leaky, in which case, the approximation described can only serve as a guideline as to the general shape of the solution curves.
\par In this vein, it is possible to provide an approximate expression in all three cases, which highlights the fact that the phase speed of the wave perturbations in the long wavelength approximation converges either to the internal cusp speed, or in a different ordering of speeds, to a value with a slight offset from this speed:
\begin{align}
\omega^2 &\approx k^2 \left[c_{T0} + f\right]^2 \left[1 + \nu (k x_0)^2\right], \qquad \text{ where } \nu>0. \label{sbgen}
\end{align}
The exact offset speed value given by $f$ depends on which band of body waves one examines, i.e.:
\begin{subequations}
\begin{alignat}{3}
f &= \max{[c_{T0}, \min{(c_1, v_{A1})}, \min{(c_2, v_{A2})} ]} - c_{T0} \quad &&\text{ for case (\ref{sba}),} \label{sbva}\\
f &= \max{[c_{T0}, \min{(c_1, v_{A1})} ]} - c_{T0} \quad &&\text{ for case (\ref{sbb}),} \label{sbvb}\\
f &= 0 \quad &&\text{ for case (\ref{sbc}).} \label{sbvc}
\end{alignat}
\end{subequations}
Substituting the appropriate form of $\omega^2$ into equation (\ref{sbthinnu}) gives us the applicable expression for $\nu$ for every (integer) $j$:
\begin{align}
\nu_j=  \frac{[(c_{T0}+f)^2-c_0^2] [v_{A0}^2 - (c_{T0} + f)^2] }{(c_0^2 + v_{A0}^2) (c_{T0} + f)^2   \pi^2 j^2 } .
\end{align}
This may then be substituted into Equation \eqref{sbgen} to obtain the approximate phase speed solutions. The corresponding quasi-kink mode may be found applying similar considerations, with the notable difference being that, here, $n_0 \cot{(n_0 x_0)}$ has to remain finite, and so $n_0 x_0 \rightarrow (j-\frac{1}{2}) \pi $ is required  (for $j=1,2,3$ ...). The values of $\nu_j$ are, in this case,
\begin{align}
\nu_j= \frac{[(c_{T0}+f)^2-c_0^2] [v_{A0}^2 - (c_{T0} + f)^2]}{(c_0^2 + v_{A0}^2) (c_{T0} + f)^2   \pi^2 (j-\frac{1}{2})^2 }  .
\end{align}
Substituting this expression back into Equation \eqref{sbgen}, it is now possible to obtain an approximation for the phase speed (and dispersion) of the slow quasi-kink body modes. Just like the quasi-sausage modes, these waves also approach the speed limit $c_{T0}+f$ bounding from below as the slab becomes thinner.

\par The fast body modes, when they exist, behave similarly to the slow body modes in the thin-slab approximation. Three bands of phase speed potentially containing body mode solutions can be distinguished:\begin{subequations}
\begin{alignat}{1}
&\max{[ \max{(c_0, v_{A0})}, \min{(c_1, v_{A1})}, \min{(c_2, v_{A2})} ]} < v_{ph}  < \min{[ \max{(c_1, v_{A1})}, \max{(c_2, v_{A2})} ]}  \label{fba}\\
&\max{[ \max{(c_0, v_{A0})}, \min{(c_1, v_{A1})}  ]} < v_{ph} < \min{[ \max{(c_1, v_{A1})}, c_{T2}  ]}  \label{fbb}\\
 &\max{(c_0, v_{A0})} < v_{ph} < \min{(  c_{T1}, c_{T2}  )} \label{fbc}.  
\end{alignat}
\end{subequations}
The question, whether the plasma-$\beta$ ($\beta=(2/\gamma)( c_0^2/ v_{A0}^2)$) is low ($c_0 <v_{A0} $) or high ($v_{A0} < c_0$), determines where the fast mode phase speeds converge to in a thin slab. Let us denote $\max{(c_0^2, v_{A0}^2)}$ with $v_{\mathrm{max}}^2$ and $\min{(c_0^2, v_{A0}^2)}$ with $v_{\mathrm{min}}^2$. Then, we may have two main cases with the same formula:
\begin{align}
\omega^2 &\approx k^2 \left[v_{\mathrm{max}} + f + u\right]^2 \left[1 + \frac{1}{\nu (k x_0)^2}\right], \qquad \text{ where } \nu>0. \label{fbgen}
\end{align}
The exact values of the lower and upper speed boundary, $f$ and $u$, depend on which band of allowed solutions one examines. In case of conditions (\ref{fba}), ..., (\ref{fbc}), we have:
\begin{subequations}
\begin{alignat}{3}
\qquad f &= \max{[ v_{\mathrm{max}}, (\min{(c_1, v_{A1})}, \min{(c_2, v_{A2})} ]} - v_{\mathrm{max}}, \label{fbva}\\
\qquad u &= \min {[ \max{(c_1, v_{A1})}, \max{(c_2, v_{A2})} ]} - f - v_{\mathrm{max}},  \nonumber \\
\qquad f &= \max{ [v_{\mathrm{max}}, (\min{(c_1, v_{A1})}  ]} - v_{\mathrm{max}},  \label{fbvb} \\
\qquad u &= \min {[ \max{(c_1, v_{A1})}, c_{T2} ]} - f - v_{\mathrm{max}} , \nonumber \\
\qquad f &=  0, \label{fbvc} \\
\qquad u &=  \min {( c_{T1}, c_{T2} )} -  v_{\mathrm{max}}, \nonumber
\end{alignat}
\end{subequations}
respectively. For the quasi-sausage modes, as before, $n_0 \tan{(n_0 x_0)}$  needs to remain finite, so $n_0 x_0$ must converge to the roots of $\tan{(n_0 x_0)} = 0$. Substituting the prescribed form of $\omega^2$ into the condition that $n_0 x_0 = j \pi $ (for $j=1,2,3$ ...) allows us to determine the possible values of $\nu$ for each of the harmonics in the direction of stratification:
\begin{align}
\nu_j=   \left\{\frac{\pi^2 j^2}{k^2 x_0^2} \frac{[v_{\mathrm{min}}^2 + v_{\mathrm{max}}^2]  [2 f v_{\mathrm{max}} + 2 u v_{\mathrm{max}} + (f + u)^2] + v_{\mathrm{max}}^4}{[(v_{\mathrm{max}} + f + u)^2 - v_{\mathrm{min}}^2] [v_{\mathrm{max}}+ f + u]^2 }-  \frac{2 f v_{\mathrm{max}} + 2 u v_{\mathrm{max}} + [f + u]^2}{[v_{\mathrm{max}} + f + u]^2} \right\}^{-1} \frac{1}{k^2 x_0^2}.
\end{align}
The quasi-kink mode under the same ordering of characteristic speeds may be shown to have coefficients of the form
\begin{align}
&\nu_j=  \left\{\frac{\pi^2 [j-\frac{1}{2}]^2}{k^2 x_0^2} \frac{[v_{\mathrm{min}}^2 + v_{\mathrm{max}}^2]  [2 f v_{\mathrm{max}} + 2 u v_{\mathrm{max}} + (f + u)^2] + v_{\mathrm{max}}^4}{[(v_{\mathrm{max}} + f + u)^2 - v_{\mathrm{min}}^2] [v_{\mathrm{max}}+ f + u]^2 } -  \frac{2 f v_{\mathrm{max}} + 2 u v_{\mathrm{max}} + [f + u]^2}{[v_{\mathrm{max}} + f + u]^2} \right\}^{-1} \frac{1}{k^2 x_0^2}.
\end{align}
Substituting these coefficients into the dispersion relation given by Equation  \eqref{fbgen} provides the approximations for the solutions of permitted wave propagation. This holds when the external sound speeds are greater than the external Alfv{\'e}n speeds. If the opposite is true, $\tan{(n_0x_0)} \rightarrow \infty$ needs to be true for quasi-sausage modes, $\cot{(n_0x_0)} \rightarrow \infty$ must hold for quasi-kink modes, and the coefficients $j$ and $j-1/2$ in the above expressions have to be modified accordingly.   \par Generally speaking, both types of fast body waves have countably many harmonics in the direction of structuring, in the phase speed band where they may exist.  It may be noted that although the effect of density ratios $\rho_0/\rho_1$ and $\rho_0/\rho_2$ cannot be seen explicitly in the calculations of this subsection, they have an indirect influence on the propagation of body waves, since they determine the values and relations of the characteristic speeds in- and outside the slab. 
\par The investigation of the thin-slab approximation has thus revealed that the introduction of magnetic asymmetry results, on the one hand, in important contributions to the dispersion of both surface- and body-mode waves and, on the other hand, in the appearance of cut-off frequencies. Beyond these frequencies, the solutions would become leaky, and therefore, when searching for trapped oscillations in the asymmetric waveguide, certain bands of phase speed must be discarded. Unlike the symmetric case, when there is one band of slow body modes, complemented by one or two bands of fast body modes, the cut-off frequencies in the asymmetric case might even result in the existence of two bands of slow mode solutions, and three bands of fast mode solutions for body waves. It can be said that, in general, the solutions are qualitatively analogue to the kink or sausage mode solutions of the symmetric case, while their exact quantitative description is more complex in the asymmetric case. Approximations can still be given for both surface- and body waves, however, thin-slab solutions for the latter will not always exist as trapped waves.

\section{Wide-slab approximation}			\label{sec:Wide} 

\par Let us now examine the waves propagating in a wide slab placed in an asymmetric magnetic environment. In solar physics, such a system could represent as an approximation of the global stratification of the atmosphere, e.g. the triad of the photosphere, the interface region, and the corona. The wide-slab approximation can also be used to model high-frequency waves present in light bridges of sunspots or elongated magnetic bright points (MBPs). 
\par In the wide-slab limit, the width of the slab is much greater than the wavelength of the waves examined, in short: $kx_0 \gg 1$. For example, only about one third of MBPs have non-circular shapes \citep{boveletmbp}, and under appropriate circumstances, they can be regarded as magnetic slabs (for details see \citeauthor{asymag} \citeyear{asymag}). These bright concentrations of magnetic flux in the photosphere are only a few hundred kilometres across \citep{solankimbp}, therefore, for any perturbations with their wavelength $ \lambda \ll 300 km $, an MBP with a width of $ 2 x_0 \approx 100 km $ can be regarded as a wide slab. For larger wavelengths, the thin slab approximation is more appropriate.
\par Light bridges between sunspot umbrae may have various widths from around 1'' up to 4'' with their extent in one direction often far greater than their length \citep{tor-15, sch-16}. A light bridge of intermediate size, with $ 2 x_0 \approx 1500 km$ (or 2'') width can then be regarded as a wide slab for waves with $ \lambda \ll 5 km$, and as a thin slab for longer wavelengths. 

\par  In the wide-slab approximation, since we have $kx_0 \gg 1$, $m_0x_0 \gg 1$ also applies (see  \citeauthor{roberts2} \citeyear{roberts2}), and so the full dispersion relation (\ref{fullsurface}) reduces to
\begin{align}
& \frac{\rho_0}{\rho_1} m_1 \frac{\rho_0}{\rho_2} m_2 \left( k^2 v_{A0}^2 - \omega^2\right)^2 +  m_0^2 \left( k^2 v_{A1}^2 - \omega^2\right) \left( k^2 v_{A2}^2 - \omega^2\right) \\  \nonumber
& \quad + \rho_0 m_0 \left( k^2 v_{A0}^2 - \omega^2\right) \left[ \frac{m_2}{\rho_2} \left( k^2 v_{A1}^2 - \omega^2\right) + \frac{m_1}{\rho_1} \left( k^2 v_{A2}^2 - \omega^2\right) \right]   =0.  
\end{align}
Since $\tanh{m_0 x_0} \rightarrow 1$ and $\coth{m_0 x_0}  \rightarrow 1$ as well, for both quasi-sausage and quasi-kink modes, the decoupled dispersion relation (\ref{surface}) takes now the same form:
\begin{align}
 (k^2 v_{A0}^2-\omega^2)  \left[ \frac{ \rho_0}{\rho_1} \frac{m_1}{ (k^2 v_{A1}^2-\omega^2) }+ \frac{ \rho_0}{\rho_2} \frac{ m_2}{ (k^2 v_{A1}^2-\omega^2)}\right] + 2   m_0= 0. \label{wideslab}
\end{align}
 \par As the width of the slab keeps increasing, the waves at one boundary will be less and less affected by the conditions at the other boundary, essentially reducing the problem to a single interface system. This may be shown by going back to the system of equations presented by the boundary conditions, namely, the continuity of velocity- and total pressure perturbation. These can be summarised in a matrix formally algebraically analogous to that in Equation (18) of \citeauthor{asymm} (\citeyear{asymm}). Rearranging the equations and substituting $\tanh{m_0 x_0} = \coth{m_0 x_0}  = 1$ into them leads to 
\begin{align}
\Lambda_i  +\Lambda_0 = 0,
\end{align} 
for $i = 1, 2$, which is the dispersion relation of a single interface (see \citeauthor{roberts1} \citeyear{roberts1}), expressed with the $\Lambda_i$ quantities defined as:
\begin{equation}
\Lambda_i = - \frac{i \rho_i}{\omega} \frac{(k^2 v_{Ai}^2 - \omega^2)}{m_i}.
\end{equation}

\par As for wide-slab body modes, the situation is similar to the thin-slab approximation, in that the results obtained for a symmetric (\citeauthor{roberts2} \citeyear{roberts2}) or asymmetric (\citeauthor{asymm} \citeyear{asymm}) slab in a non-magnetic environment can be generalised, so that the constraints set by the external densities and magnetic fields will now also be taken into account. The phase speed of slow body modes, which would converge to $v_{\mathrm{min}}$ in a field-free environment, might only do so with some offset, which may be described as
\begin{align}
\omega^2 = k^2 [v_{\mathrm{min}}- u]^2 \left[ 1 + \frac{\nu}{(k x_0)^2} \right] , \label{stypeawide}
\end{align}
where the exact value of $u$ depends on which band of solutions we examine, i.e. in case of (\ref{sba})-(\ref{sbc}) \begin{subequations}
\begin{alignat}{3}
u&= v_{\mathrm{min}} - \min{[ \min{(c_0, v_{A0})}, \max{(c_1, v_{A1})} , \max{(c_2, v_{A2})}  ]},\\
u&= v_{\mathrm{min}} - \min{[ \min{(c_0, v_{A0})}, \max{(c_1, v_{A1})} , c_{T2}  ]}, \\
u&= v_{\mathrm{min}} - \min{[ \min{(c_0, v_{A0})}, c_{T1}, c_{T2}  ]}, 
\end{alignat}
\end{subequations}
respectively. For the quasi-sausage mode solutions, as $kx_0 \rightarrow \infty$, the condition can be set that $\tan{(n_0 x_0)} \rightarrow \pm \infty$, which means for the argument that $n_0 x_0 \rightarrow (j-\frac{1}{2}) \pi$. This gives us the $\nu_j$ coefficients as  
\begin{align}
\nu_j=  \pi^2 \left[j-\frac{1}{2}\right]^2 \frac{[v_{\mathrm{min}}^4 - (v_{\mathrm{min}}^2 + v_{\mathrm{max}}^2) (2 u v_{\mathrm{min}} - u^2)]}{[v_{\mathrm{max}}^2 - (v_{\mathrm{min}} - u)^2] [v_{\mathrm{min}}- u]^2  }  . \label{stypeanusaus}
\end{align}
\par The slow body quasi-kink modes can be found in a similar fashion, by setting  $n_0 x_0 \rightarrow j \pi$ so that $\cot{(n_0 x_0)} \rightarrow \pm \infty$. This leads to
\begin{align}
&\nu_j= \pi^2 j^2 \frac{[v_{\mathrm{min}}^4 - (v_{\mathrm{min}}^2 + v_{\mathrm{max}}^2) (2 u v_{\mathrm{min}} - u^2)]}{[v_{\mathrm{max}}^2 - (v_{\mathrm{min}} - u)^2] [v_{\mathrm{min}}- u]^2  }. \label{stypeanukink}
\end{align}
Substituting these into Equation \eqref{stypeawide} gives us the approximations of the body modes in a wide slab. This holds when $v_{A0} > c_0$. In a high-beta slab, however, the condition for the quasi-sausage modes becomes $\tan{(n_0x_0)} \rightarrow 0$, while for quasi-saisage modes, $\cot{(n_0x_0)} \rightarrow 0$, and the expressions containing the coefficients $j$ and $j-1/2$ have to be adjusted accordingly. 
\par An analogous derivation leads to the approximate solutions for fast mode body waves in the wide slab. These modes can be assumed to tend towards the higher internal characteristic speed in the field-free configuration in the limit of short wavelength approximation. In the magnetically asymmetric configuration, their dispersion is expected to follow
\begin{align}
&\omega^2 =  k^2 [v_{\mathrm{max}} + f]^2  \left[1 +  \frac{1}{(k x_0)^2 \nu } \right], \label{fastwide}
\end{align}
where the exact value of $f$ depends on which band of solutions one takes. In case (\ref{fba}), (\ref{fbb}) and (\ref{fbc}), the factors $f$ are defined by Equations \eqref{fbva}, \eqref{fbvb} and \eqref{fbvc}, respectively. 
\par For quasi-sausage modes, $n_0 x_0 \rightarrow (j-\frac{1}{2}) \pi$ has to be true for $n_0 \tan{(n_0 x_0)}$ to remain finite. This leads to
\begin{align}
&\nu_j=  \left\{  \pi^2 \left[j-\frac{1}{2}\right]^2 \frac{ [(v_{\mathrm{min}}^2 + v_{\mathrm{max}}^2) (2 f v_{\mathrm{max}} + f^2) + v_{\mathrm{max}}^4] }{[(v_{\mathrm{max}} + f)^2 - v_{\mathrm{min}}^2] [v_{\mathrm{max}} + f]^2 }  - \frac{[2 f v_{\mathrm{max}} + f^2] [k x_0]^2 }{[v_{\mathrm{max}} + f]^2} \right\}^{-1}. \label{fwnusaus} 
\end{align}
Similarly, for the quasi-kink modes, $n_0 x_0 \rightarrow j \pi$, so the coefficients and the frequencies are only marginally different:
\begin{align}
&\nu_j=  \left\{ \pi^2 j^2 \frac{[(v_{\mathrm{min}}^2 + v_{\mathrm{max}}^2) (2 f v_{\mathrm{max}} + f^2) + v_{\mathrm{max}}^4] }{[(v_{\mathrm{max}} + f)^2 - v_{\mathrm{min}}^2] [v_{\mathrm{max}} + f]^2 } - \frac{[2 f v_{\mathrm{max}} + f^2] [k x_0]^2}{[v_{\mathrm{max}} + f]^2} \right\}^{-1}. \label{fwnukink}
\end{align}
Substituting the appropriate coefficient $\nu$ from Equations \eqref{fwnusaus} and \eqref{fwnukink}, respectively, into Equation \eqref{fastwide} gives us the quasi-sausage and quasi-kink mode solutions for the MHD wave propagation in the wide-slab approximation. This is true when $c_0 > v_{A0}$. In a low-beta slab, however, the condition for quasi-sausage modes is $\tan{(n_0 x_0)} \rightarrow 0$, while for quasi-kink modes, it is $\cot{(n_0 x_0)} \rightarrow 0$. Further, the coefficients $j$ and $j-1/2$ in the above expressions have to be swapped to fulfil these conditions.
\par Much like in the thin-slab approximation, the effect of the differences in the equilibrium parameters in the external environment on body modes is not obvious immediately. To second order, there are no terms containing the density ratios, unlike for the surface waves. 
Overall, we may conclude that a magnetically asymmetric environment has greater effect on MHD surface waves than on body modes. Applications to solar and astrophysical plasmas may be exploited, e.g. by means of solar magneto-seismology. Such analysis may be performed with greater success for MHD waves observed in magnetic structures that can be modelled by the thin-slab approximation, since in wide slabs, the effects of asymmetry can be felt to a lesser degree at either of the interfaces, which are distant from each other.

\section{Low-$\beta$ approximation}		\label{sec:Lowbeta} 

\par In the low-$\beta$ approximation, the magnetic pressure dominates the gas pressure in a given region of plasma ($\beta_i = p_{i}/p_{i,m} << 1$, for $i=0,1$ or $2$.) Therefore in the low-$\beta$ limit, $c_i/v_{Ai} <<1$. This particular approximation has practical as well as analytical use: it reduces the dispersion relation into a simpler form, and it also has a very significant range of applicability, since from about the mid-chromosphere upwards into the corona, the solar atmosphere is considered to be a low-$\beta$ environment. This is exactly the case that we are first going to investigate in the following section, using a model in which the plasma-$\beta$ is low in all three domains. Afterwards, we will describe the limiting case, whereby all three domains of the asymmetric slab system are filled with cold plasma (that is, $\beta_i =0$, for $i=0,1,2$). This considerably simplifies the analytical expressions describing wave dispersion, while it still approximates well the low values of plasma-$\beta$ found in upper solar atmospheric, e.g., in coronal conditions.

\subsection{Low plasma-$\beta$ in all three domains}		\label{sec:lll} 

\par In the case when the plasma-$\beta$ is low, but non-zero, it is possible to express the coefficients $m_0$, $m_1$, $m_2$ in terms of $\beta_0, \beta_1, \beta_2$, and apply some simplifications to the dispersion relation. This way, the modified wavenumber coefficients become
	\begin{align}
		{m}_{{i}}^{{2}}&=\frac{\left({{k}^{{2}}   \beta_i \gamma  {v}_{{Ai}}^{{2}}- 2 {\omega }^{{2}}}\right)\left({{k}^{{2}}{v}_{{Ai}}^{{2}}-{\omega}^{{2}}}\right)}{\left({{k}^{{2}}  \beta_i \gamma {v}_{Ai}^{{4}}-  \beta_i \gamma {v}_{Ai}^{{2}}{\omega }^{{2}}- 2 v_{{Ai}}^{{2}}{\omega }^{{2}}}\right)},  \qquad \text{for } i=0,1,2, \label{lowcoeff2}\\
		{n}_{{0}}^{{2}}&=\frac{\left({{k}^{{2}}   \beta_i \gamma  {v}_{{Ai}}^{{2}}- 2 {\omega }^{{2}}}\right)\left({{\omega}^{{2}} - {k}^{{2}}{v}_{{Ai}}^{{2}}}\right)}{\left({{k}^{{2}}  \beta_i \gamma {v}_{Ai}^{{4}}-  \beta_i \gamma {v}_{Ai}^{{2}}{\omega }^{{2}}- 2 v_{{Ai}}^{{2}}{\omega }^{{2}}}\right)}.  \label{lowcoeff1}
	\end{align}
Assuming the plasma-$\beta$ is small in all three domains, an expansion of the dispersion relation about $(\beta_0, \beta_1, \beta_2) \approx (0, 0, 0)$ can be performed. Taking only zeroth- and first-order terms into consideration, the dispersion relation for surface modes takes the following form:
\begin{align}
L_1 &+ L_2 + L_{0s} - \frac{\gamma}{4} \left\{ L_1 \beta_1 + L_2 \beta_2 + L_{0s} \beta_0  \pm \frac{2 x_0 \beta_0 }{v_{A0}^2} \left[ 1 - \binom{\mathrm{tanh}^2}{\mathrm{coth}^2}\{m_{0z} x_0\} \right] \right\}=0,
\end{align}	
where
\begin{align}
L_j &= \frac{\rho_0}{\rho_j} \frac{m_{jz}}{(k^2 v_{Aj}^2 - \omega^2)}, \qquad \text{for } j=1,2, \\
L_{0s} &= \frac{2 m_{0z}}{(k^2 v_{A0}^2 - \omega^2)} \binom{\tanh{}}{\coth{}} \{m_{0z} x_0\}, \\
m_{iz} &= \left(  \frac{k^2 v_{Ai}^2 - \omega^2}{v_{Ai}^2} \right)^{1/2}, \qquad \text{for } i=0,1,2. \label{lm0z}
\end{align}	
Here, the index 'z' denotes the form of the wavenumber coefficients when $\beta=0$ in the given domain, and the index 's' refers to the fact that the term $L_{0s}$ in necessary for the description of \textit{surface} waves. In this term, the parts containing the $\tanh{}$, $\coth{}$ functions describe quasi-sausage and quasi-kink \textit{surface} modes, respectively. With the same notation, the expansion of the dispersion relation for \textit{body} waves becomes
\begin{align}
L_1 &+ L_2 + L_{0b} - \frac{\gamma}{4} \left\{ L_1 \beta_1 + L_2 \beta_2 +  \beta_0 \left[ L_{0b} \mp \frac{1}{2} L_{0b}^2 x_0  (k^2 v_{A0}^2 - \omega^2) \mp \frac{2 n_{0z}^2 x_0}{(k^2 v_{A0}^2 - \omega^2)}\right] \right\}=0,
\end{align}	
where further
\begin{align}
L_{0b} &= \frac{2 n_{0z}}{(k^2 v_{A0}^2 - \omega^2)} \binom{-\tan{}}{\cot{}} \{n_{0z} x_0\}, \\
n_{0z} &= \left(  \frac{ \omega^2 - k^2 v_{Ai}^2}{v_{Ai}^2} \right)^{1/2}. \label{ln0z}
\end{align}	
Here, the index 'b' expresses that the term $L_{0b}$ is required for the description of body modes, and, again, the upper part (with the $\tan{}$ function and the minus signs) describes quasi-sausage \textit{body} modes, while the lower part governs the dispersion of the quasi-kink \textit{body} modes.

   \begin{figure}   

   \centerline{\hspace*{0.001\textwidth}
               \includegraphics[width=0.425\textwidth,height=0.4\textwidth,keepaspectratio]{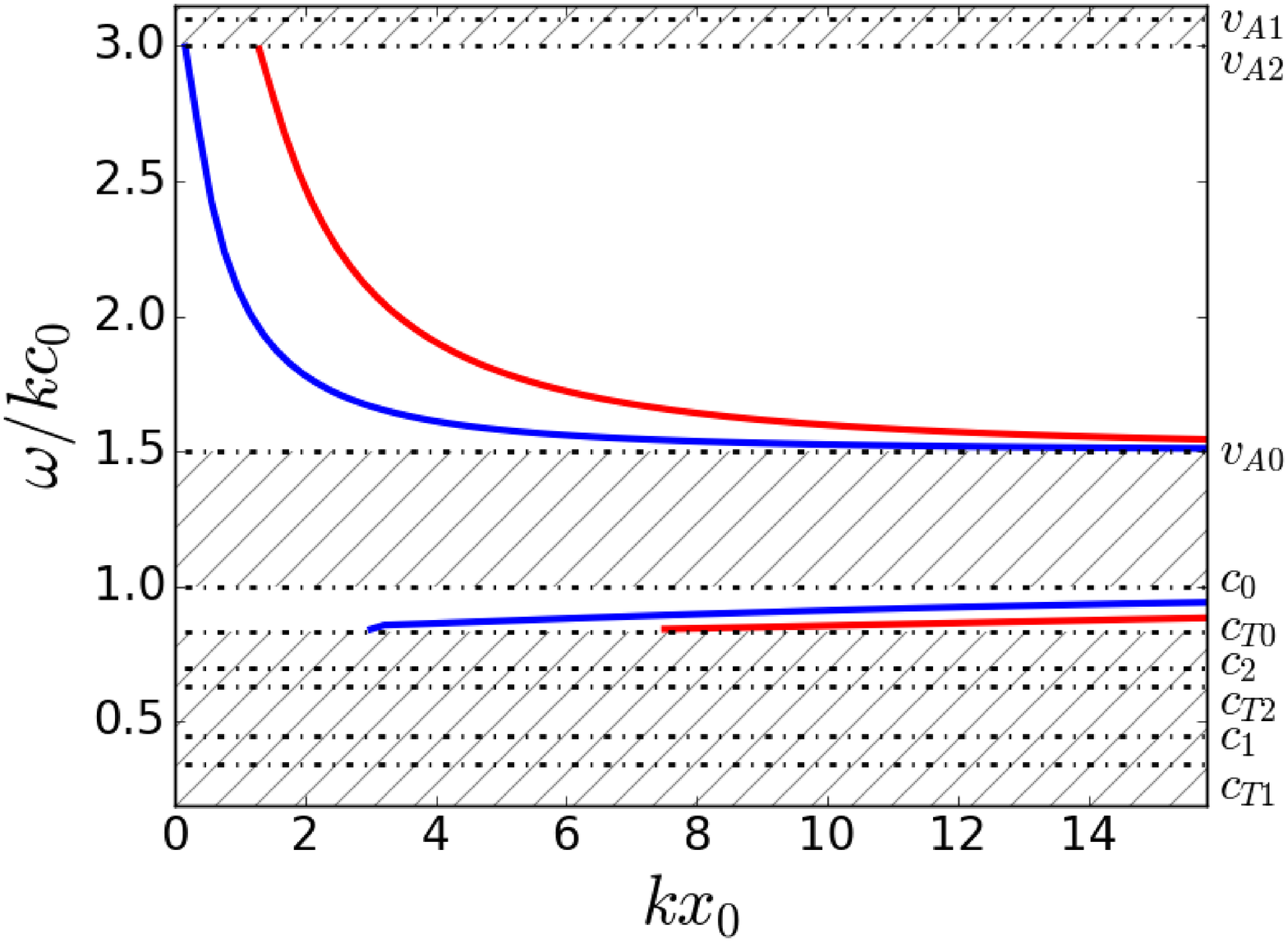}
               \hspace*{0.02\textwidth}
               \includegraphics[width=0.425\textwidth,height=0.4\textwidth,keepaspectratio]{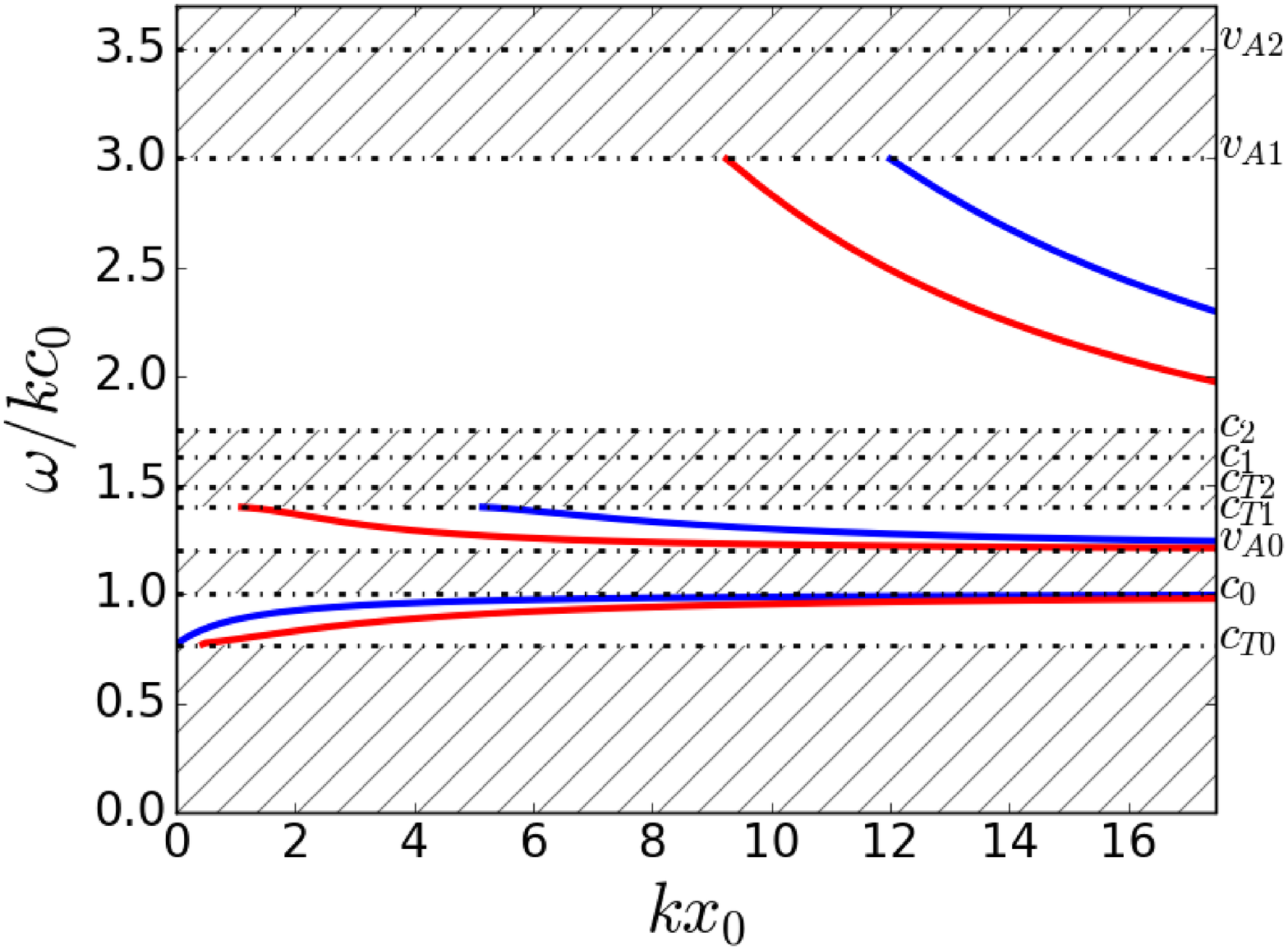}
              }
     \vspace{-0.3\textwidth}  
     \centerline{\Large \bf      
      \hspace{0.04 \textwidth}  \color{black}{(a)}
      \hspace{0.4\textwidth}  \color{black}{(b)}
         \hfill}
     \vspace{0.28\textwidth} 

   \centerline{\hspace*{0.001\textwidth}
               \includegraphics[width=0.425\textwidth,height=0.4\textwidth,keepaspectratio]{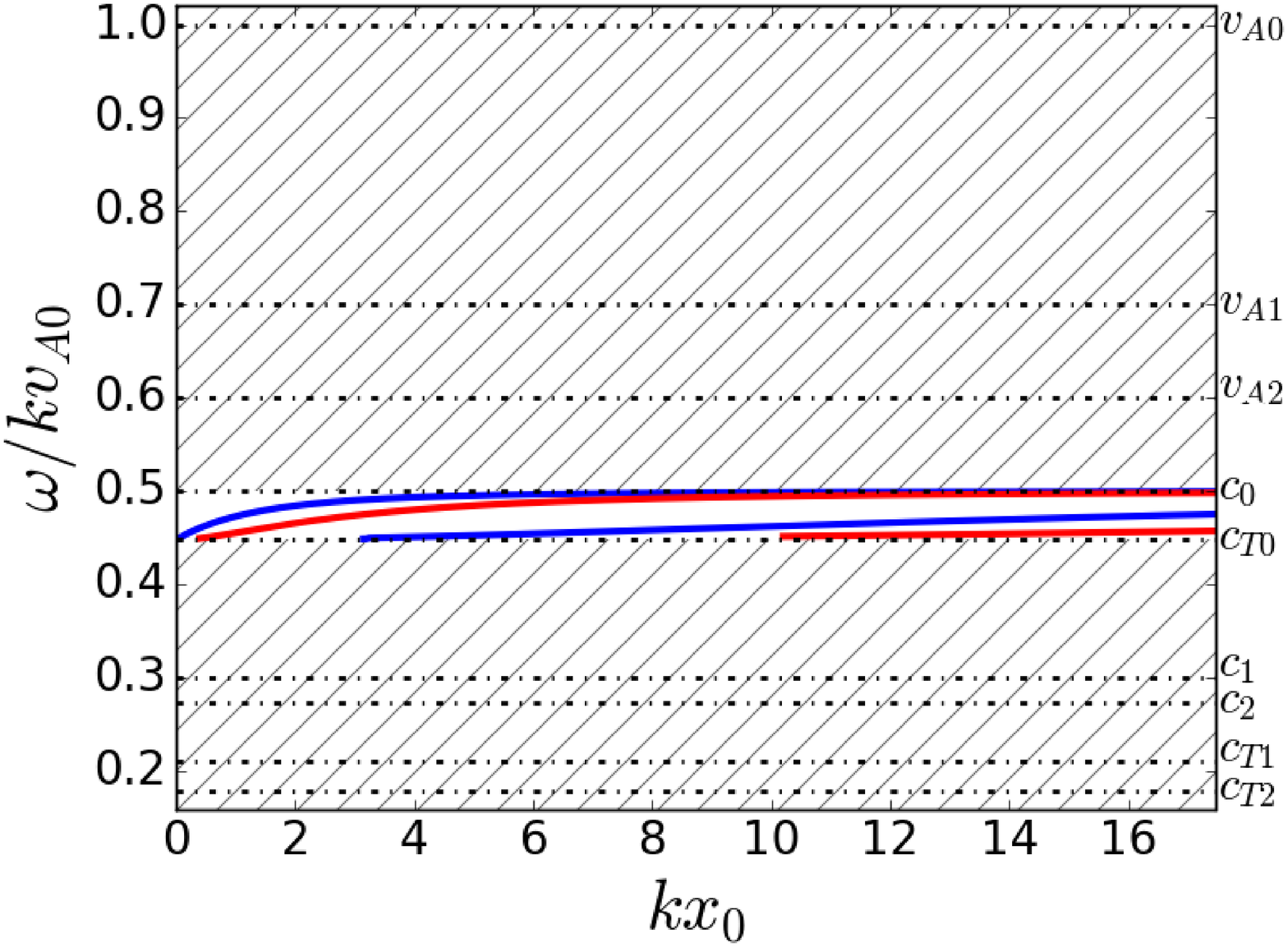}
               \hspace*{0.02\textwidth}
               \includegraphics[width=0.425\textwidth,height=0.4\textwidth,keepaspectratio]{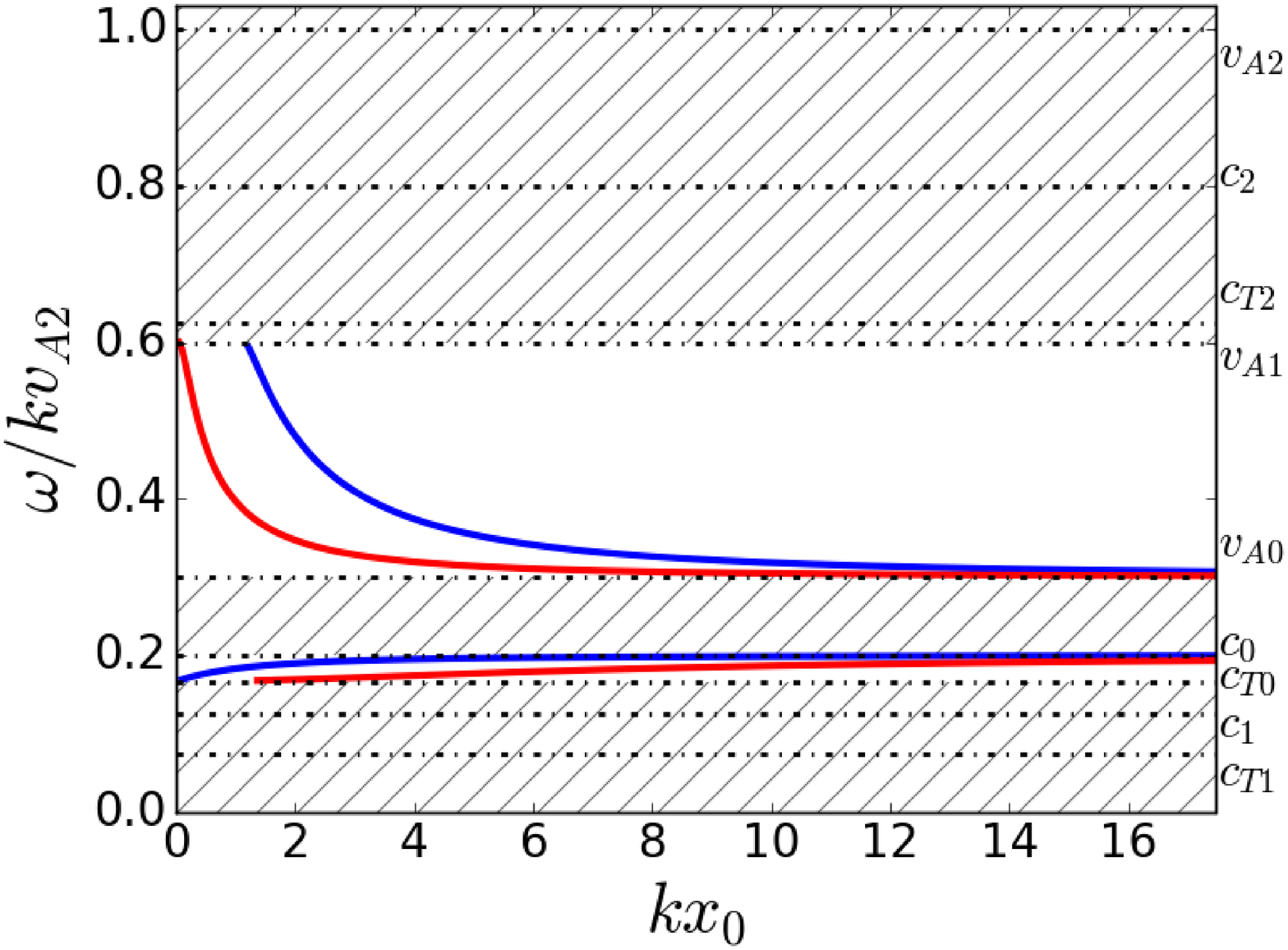}
              }
     \vspace{-0.3\textwidth}  
     \centerline{\Large \bf     
      \hspace{0.04 \textwidth} \color{black}{(c)}
      \hspace{0.4\textwidth}  \color{black}{(d)}
         \hfill}
     \vspace{0.28\textwidth}    
\caption{The phase speed ($\omega/k$) of magnetoacoustic waves that occur in various low-$\beta$ situations characterised by typical choices of $c_i, v_{Ai}, \rho_i$. Blue (red) curves show quasi-sausage (quasi-kink) modes. Hatching represents regions in which no propagating modes are permitted. \textbf{(a)} Slow and fast mode body waves are visualised when $v_{A0}=1.5 c_0$, $v_{A1}=4 c_0$, $v_{A2}=3 c_0$, $c_{1}=0.5976 c_0$, $c_{2}=0.6972 c_0$, $\rho_1/\rho_0=0.21$, $\rho_2/\rho_0=0.36$. \textbf{(b)} One band of slow-, and two bands of fast body modes appear when  $v_{A0}=1.2 c_0$, $v_{A1}=3 c_0$, $v_{A2}=3.5 c_0$, $c_{1}=1.5811 c_0$, $c_{2}=1.6531 c_0$, $\rho_1/\rho_0=0.22$, $\rho_2/\rho_0=0.17$. \textbf{(c)} Only slow body modes can be found when $v_{A1}=0.7 v_{A0}$, $v_{A2}=0.6 v_{A0}$, $c_0= 0.5 v_{A0}$, $c_{1}=0.2504 v_{A0}$, $c_{2}=0.2472 v_{A0}$, $\rho_1/\rho_0=2.3$, $\rho_2/\rho_0=3.0$. \textbf{(d)} Even with more prominent asymmetry, one band of slow-, and one band of fast body modes exist when, e.g., $v_{A0}=0.3 v_{A2}$, $v_{A1}=0.6 v_{A2}$, $c_0= 0.2 v_{A2}$, $c_{1}=0.1 v_{A2}$, $c_{2}=0.8 v_{A2}$, $\rho_1/\rho_0=0.3710$, $\rho_2/\rho_0=0.0871$. In each panel, only a couple of examples in each band of body modes are displayed.}
   \label{fig:lowbeta}              
   \end{figure}

\par \citeauthor{roberts3} (\citeyear{roberts3}) explored the low-$\beta$ case in a magnetic, but symmetric environment of the slab, and their qualitative basic findings still hold in an asymmetric slab. Let us now solve the dispersion relation for a few different and representative asymmetric slab systems filled in all three domains with low-$\beta$ plasma, and visualise the wave spectrum.  Panel (a) of Figure (\ref{fig:lowbeta}) shows that, when the ordering of the characteristic propagation speeds is $c_i < c_0 < v_{A0} < v_{Ai}$ (where $i=1,2$), no surface modes, but only body waves can be found. The slow body waves have phase speed $ c_{T0} < v_{ph} < c_0$, and the fast body waves propagate with $v_{A0} < v_{ph} < v_{A2}$, which corresponds to the conditions outlined in case (\ref{sba}) for the slow waves, and case (\ref{fba}) for the fast waves. Both the quasi-sausage and the quasi-kink modes are present. 
\par If the slab is cooler than its environment (the sound speeds are interchanged, so the ordering is $c_0 < c_e < v_{A0} < v_{Ai}$), the result is similar: both fast and slow waves may be present, as panel (b) of Figure (\ref{fig:lowbeta}) illustrates. A further interesting observation can be made in this equilibrium configuration. While the slow modes represent case (\ref{sbc}), there are two bands of body waves, corresponding to the conditions in (\ref{fba}) for the faster band, and (\ref{fbc}) for the slower band. A similar result was obtained by \citeauthor{roberts3} (\citeyear{roberts3}) for the symmetric case, illustrated in their Figure 7.  
\par The situation is vastly different, however, if the Alfv{\'e}n speeds are interchanged (compared to the original ordering shown in panel (a)). In this case, presented in panel (c) of Figure (\ref{fig:lowbeta}), the internal Alfv{\'e}n speed is higher than both external Alfv{\'e}n speeds, and, just as in the symmetric slab, only the slow body waves remain possible (corresponding to the conditions in (\ref{sba})).
\par Panel (d) of Figure (\ref{fig:lowbeta}) demonstrates that even if the asymmetry is great enough in the system so that the internal sound speed falls between the external ones, two bands of body modes remain possible. The slow band is defined by the criteria of (\ref{sbb}), with phase speeds falling between $c_{T0} < v_{ph} <c_0$. The band of fast body waves possess phase speeds in the range $v_{A0} < v_{ph} < v_{A1}$, corresponding to case (\ref{fbb}).

\subsection{Zero-$\beta$ limit}		\label{sec:Zerobeta} 

\par An extreme but often practical case of the low-$\beta$ approximation is the zero-$\beta$ limit, in which the sound speeds are negligible as compared to the Alfv{\'e}n speeds: $c_1 \approx c_2 \approx c_0 \approx 0$, which can be said to describe coronal plasma conditions using the MHD framework. This assumption also leads to a vastly simplified equation for the description of wave dispersion. The zero-$\beta$ approximation eliminates slow body waves, and only the fast body waves remain possible, just like in the symmetric case \citep{roberts3}.
\par In the zero-$\beta$ limit the modified wavenumber coefficients are given by Equations \eqref{lm0z} and \eqref{ln0z}, and the first-order terms of the expanded dispersion relation vanish, leaving
\begin{align}
\binom{\tan{}}{-\cot{}} \{n_{0z} x_0\}  = \frac{1}{2} \frac{\rho_0}{\rho_1} \frac{v_{A0} (k^2 v_{A0}^2 - \omega^2)^{1/2}} {v_{A1} (k^2 v_{A1}^2 - \omega^2)^{1/2}} + \frac{1}{2}  \frac{\rho_0}{\rho_2} \frac{v_{A0} (k^2 v_{A0}^2 - \omega^2)^{1/2}}{v_{A2} (k^2 v_{A2}^2 - \omega^2)^{1/2}}  ,
\end{align}	
Total pressure balance must be upheld at both interfaces of the asymmetric slab. In terms of the characteristic speeds, this condition can be expressed as
\begin{align}
	\frac{\rho_i}{\rho_j}= \frac{c_j^2 + \frac{1}{2} \gamma v_{Aj}^2}{c_i^2 + \frac{1}{2} \gamma v_{Ai}^2}, \quad \text{ where } i=0,1,2; \quad j=0,1,2; \quad i \neq j. \label{eq:ratio}
\end{align}
Since the sound speeds are zero in this limit, Equation \eqref{eq:ratio} can be used to further simplify the dispersion relation:
\begin{align}
 \binom{\tan}{-\cot} \{n_{0z} x_0\}  &= -\frac{1}{2} \left( \frac{n_{0z}}{m_{1z}} + \frac{n_{0z}}{m_{2z}} \right). \label{zerobeta} 
\end{align}
In the fully symmetric case, this expression reduces to Equations (22) and (23) of \citeauthor{roberts3} (\citeyear{roberts3}).
\par In Equation \eqref{zerobeta}, $n_{0z}$, $m_{1z}$ and $m_{2z}$ $>$ $0$, which are only true when $k^2 v_{A0}^2 < \omega^2 < \min{(k^2 v_{A1}^2, k^2 v_{A2}^2)}$. The role of asymmetry manifests in this selection for the lower Alfv{\'e}n speed value. An alternate description of body waves in this band, e.g. in the wide-slab limit, can be constructed by the substitution of $\omega^2 = k^2 v_{A, \mathrm{min}}^2 [\rho_{ \mathrm{min}}/\rho_0] \left[ 1 + \nu/(kx_0)^2 \right]$, where the index $m$ denotes external equilibrium parameters of the side with the lower (external) Alfv{\'e}n speed. Applying the same considerations that we used while deriving the wide-slab approximation in the general case allows us to determine the coefficients $\nu_j$. This process yields the expression 
\begin{align}
\omega^2 = k^2 v_{A, \mathrm{min}}^2 \frac{\rho_{\mathrm{min}}}{\rho_0} \left[ 1 + \frac{\pi^2 \left( j - \frac{1}{2}\right)^2}{k^2 x_0^2} \right] \label{zerobetas}
\end{align}
for quasi-sausage modes, and
\begin{align}
\omega^2   = k^2 v_{A, \mathrm{min}}^2 \frac{\rho_{\mathrm{min}}}{\rho_0} \left[ 1 + \frac{\pi^2  j^2}{k^2 x_0^2} \right] \label{zerobetak}
\end{align}
for quasi-kink modes of the fast body wave. A basic diagnostic purpose may be fulfilled by making these approximations. Namely, for given values of $j$, $\omega$ and $k$, Equations \eqref{zerobetas} and \eqref{zerobetak} determine a simple connection between the lower external Alfv{\'e}n speed and the external-to-internal density ratio on the same side, therefore, knowing one of them can provide an estimate of the other.
The description of eigenmodes  in the low- and zero-$\beta$ asymmetric slab is formally analogous to that in the symmetric case. However, the difference in equilibrium external parameters - even in this simplified scenario - adds some analytical complexity. Perhaps the most important difference resulting from the asymmetry is that, although the fast body mode solution curves are still located between the external and internal Alfv{\'e}n speeds, they experience a cut-off in the thin-slab limit: with phase speed above the lower external Alfv{\'e}n speed, the waves become leaky.

\section{High-$\beta$ approximation}		\label{sec:Highbeta} 

In the approximation of high plasma-$\beta$ magnetic pressure is dominated by plasma kinetic pressure. Since this is more generally true for lower solar atmospheric conditions, it is worthwhile to explore the behaviour of wave perturbations in this limit of plasma and magnetic parameters. First, we are going to derive the dispersion relation for the case of high plasma-$\beta$ in all three domains, and provide examples of its numerical solution. Further on, we will demonstrate the analytical ease that the extreme infinite-$\beta$ approximation brings to the problem.

\subsection{High plasma-$\beta$ in all three domains} 		\label{sec:hhh} 

\par If the plasma-$\beta$ is high, the Alfv{\'e}n speeds are negligible compared to the sound speeds of each domain: $c_{i}/v_{Ai} \gg 1$ for $i=0,1,2$. In this limit, the modified wavenumber coefficients take the following form:
	\begin{align}
		{m}_{{i}}^2&=\frac {({{k}^{{2}}{c}_{{i }}^{{2}}-{\omega }^{{2}}}) ({2 {k}^{{2}}  {c}_{{i}}^{{2}}- \gamma {\beta }_{{i}} {\omega }^{{2}}})}{{c}_{{i}}^2 ({2 {k}^{{2}} {c}_{{i}}^{{2}}- 2 {\omega }^{{2}}- \gamma {\beta }_{{i}} {\omega }^{{2}}})} \quad \text{ for } i=0,1,2, \label{highcoeff2} \\
		{n}_{{0}}^2&=\frac {({{\omega }^{{2}}- {k}^{{2}}{c}_{{0 }}^{{2}}}) ({2 {k}^{{2}}  {c}_{{0}}^{{2}}- \gamma {\beta }_{{0}} {\omega }^{{2}}})}{{c}_{{0}}^2 ({2 {k}^{{2}}   {c}_{{0}}^{{2}}- 2   {\omega }^{{2}}- \gamma {\beta }_{{0}} {\omega }^{{2}}})}. \label{highcoeff1}
	\end{align}
\par
\begin{figure}    
                                 
   \centerline{\hspace*{0.001\textwidth}
               \includegraphics[width=0.43\textwidth,height=0.4\textwidth,keepaspectratio]{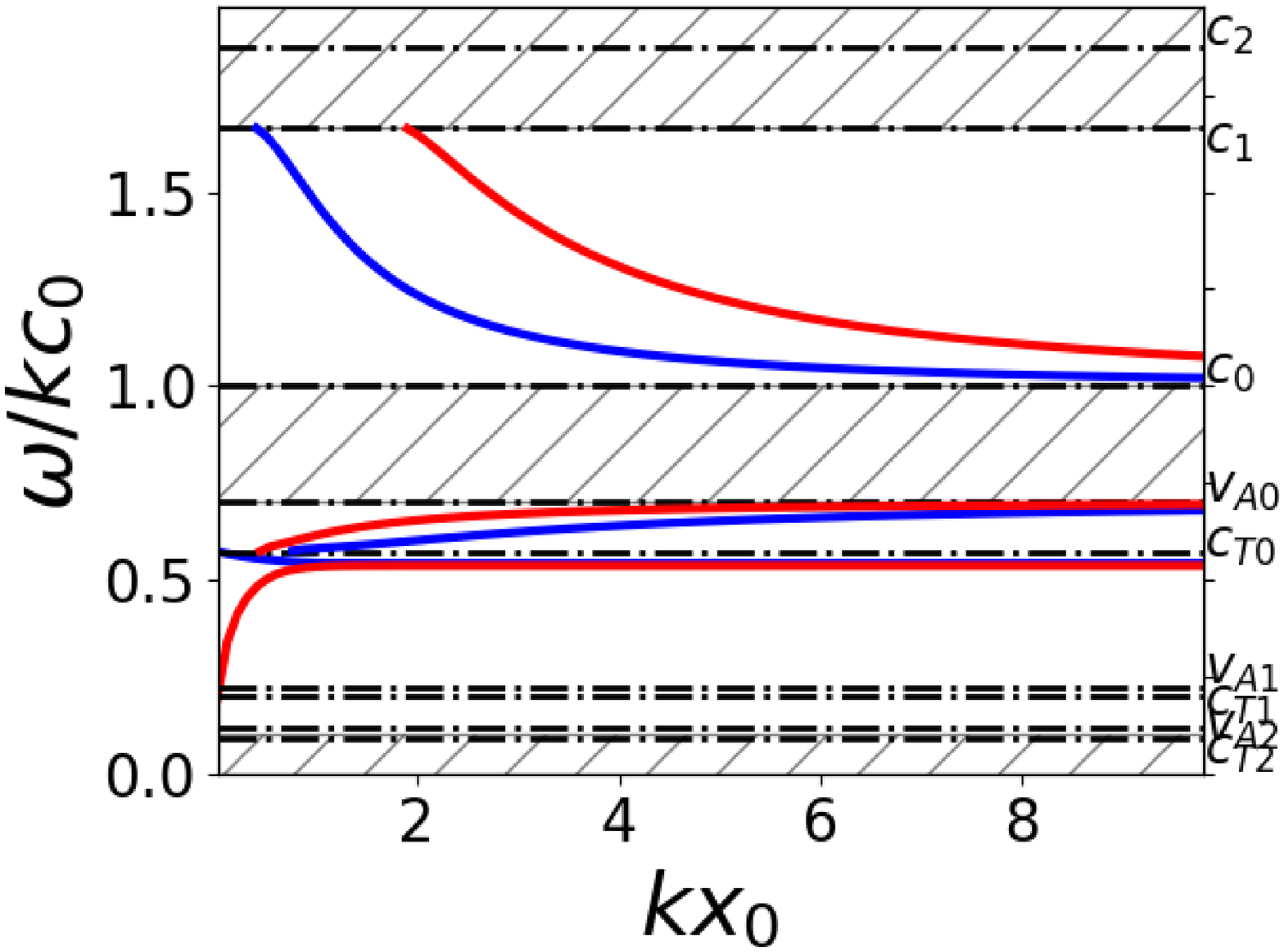}
               \hspace*{0.02\textwidth}
               \includegraphics[width=0.425\textwidth,height=0.4\textwidth,keepaspectratio]{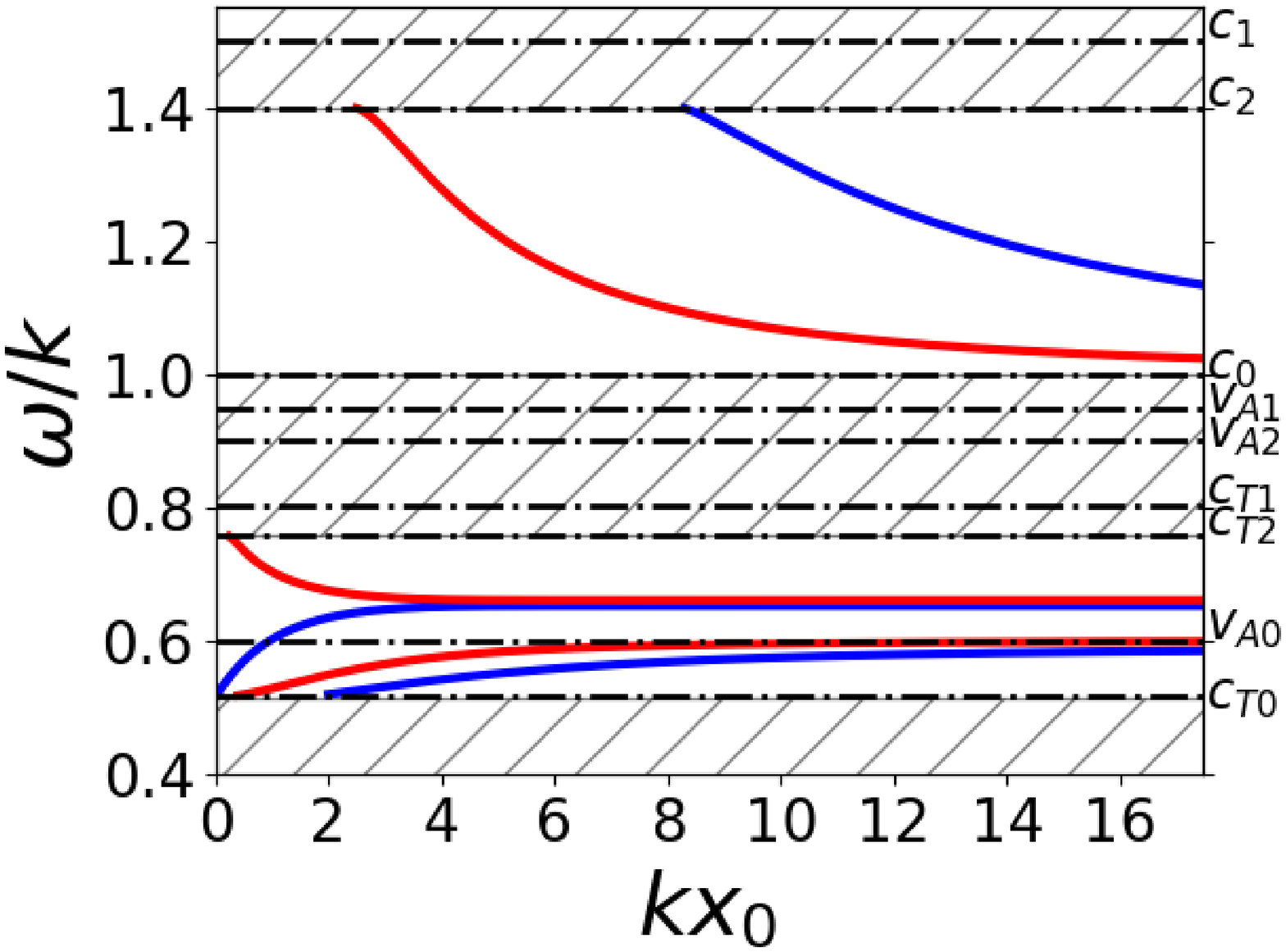}
              }
     \vspace{-0.3\textwidth}  
     \centerline{\Large \bf      
      \hspace{0.04 \textwidth}  \color{black}{(a)}
      \hspace{0.4\textwidth}  \color{black}{(b)}
         \hfill}
     \vspace{0.28\textwidth} 

   \centerline{\hspace*{0.001\textwidth}
               \includegraphics[width=0.43\textwidth,height=0.4\textwidth,keepaspectratio]{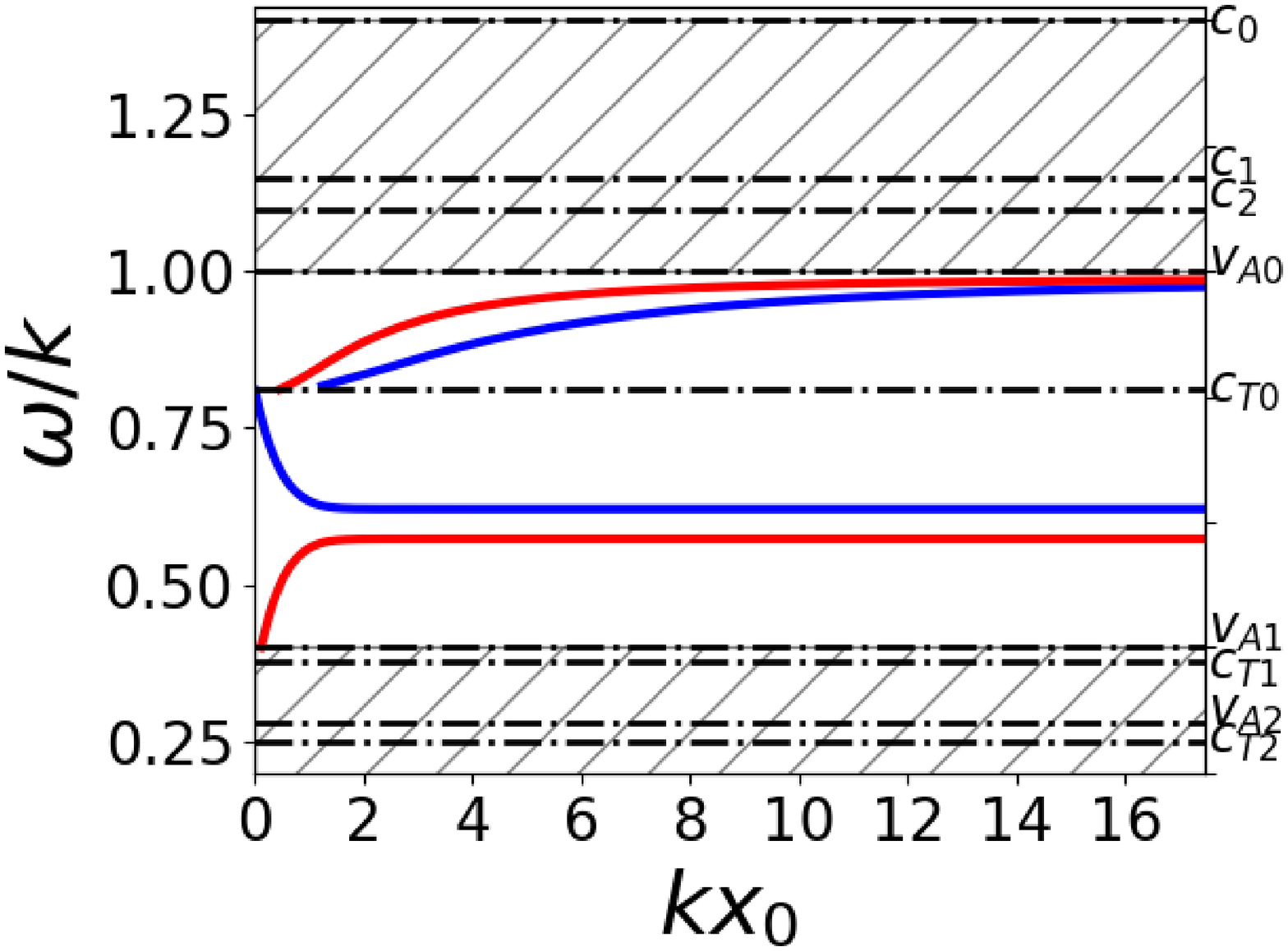}
               \hspace*{0.02\textwidth}
               \includegraphics[width=0.425\textwidth,height=0.4\textwidth,keepaspectratio]{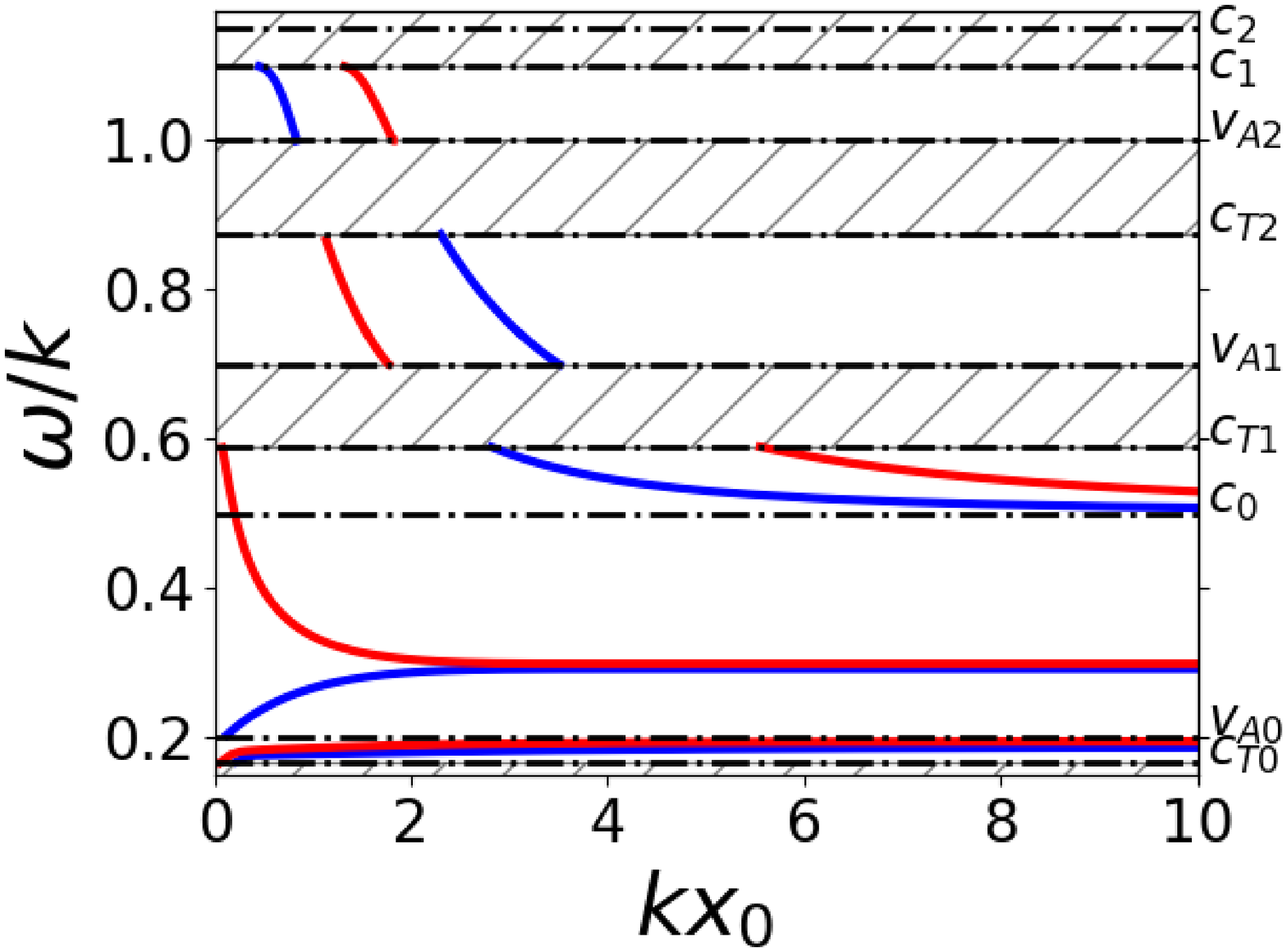}
              }
     \vspace{-0.3\textwidth}  
     \centerline{\Large \bf     
      \hspace{0.04 \textwidth} \color{black}{(c)}
      \hspace{0.4\textwidth}  \color{black}{(d)}
         \hfill}
     \vspace{0.28\textwidth}    
\caption{Solutions to the dispersion relation, similar as Figure \ref{fig:lowbeta}, but for high-$\beta$ cases. \textbf{(a)} Slow and fast mode body waves, as well as slow surface waves are present when $v_{A0}=0.7 c_0$, $v_{A1}=0.2 c_0$, $v_{A2}=0.1 c_0$, $c_{1}=1.6683 c_0$, $c_{2}=1.8742 c_0$, $\rho_1/\rho_0=0.5$, $\rho_2/\rho_0=0.4$. \textbf{(b)} The same modes appear when  $v_{A0}=0.6 c_0$, $v_{A1}=0.95 c_0$, $v_{A2}=0.9 c_0$, $c_{1}=1.5 c_0$, $c_{2}=1.4 c_0$, $\rho_1/\rho_0=0.433$, $\rho_2/\rho_0=0.4934$. \textbf{(c)} Only slow surface and body modes can be observed when $v_{A1}=0.4 v_{A0}$, $v_{A2}=0.3 v_{A0}$, $c_0= 1.4 v_{A0}$, $c_{1}=1.15 v_{A0}$, $c_{2}=1.1 v_{A0}$, $\rho_1/\rho_0=1.9188$, $\rho_2/\rho_0=2.1738$. \textbf{(d)} Three bands of fast body modes, one band of slow body modes, and a pair of slow surface modes exist when  $v_{A0}=0.2 v_{A2}$, $v_{A1}=0.7 v_{A2}$, $c_0= 0.5 v_{A2}$, $c_{1}=1.1 v_{A2}$, $c_{2}=1.8 v_{A2}$, $\rho_1/\rho_0=0.1751$, $\rho_2/\rho_0=0.071$. In each panel, only a couple of examples in each band of body modes are displayed.}
   \label{fig:hhh}              
   \end{figure}

 Several modes are possible in this case, which is illustrated in Figure \ref{fig:hhh}, including both surface and body waves. For an analytical description of the wave modes, the dispersion relation can be expanded about $(1/\beta_0, 1/\beta_1, 1/\beta_2) \approx (0,0,0)$. Keeping only zeroth- and first-order terms then yields
\begin{align}
H_1 &+ H_2 + H_{0s} + \frac{1}{\gamma \omega^2} \left\{  \left[ 2 k^2 c_1^2 - \omega^2  \right] \frac{H_1}{\beta_1} +  \left[ 2 k^2 c_2^2 - \omega^2  \right] \frac{H_2}{\beta_2} \right.  \nonumber \\
& \quad \left. +   \left[ 2 k^2 c_0^2 - \omega^2  \right] \frac{H_{0s}}{\beta_0}  +  \frac{2 x_0 m_{0z}^2}{\beta_0} \left[ 1 - \binom{\mathrm{tanh}^2}{\mathrm{coth}^2}\{ m_{0z} x_0\} \right] \right\} =0
\end{align}
for surface waves, where
\begin{align}
H_j &= - \frac{\rho_0}{\rho_j} \frac{m_{jz}}{\omega^2}, \qquad \text{for } j=1,2, \\
H_{0s} &= - \frac{2 m_{0z}}{ \omega^2 } \binom{\tanh{}}{\coth{}} \{m_{0z} x_0\}, \\
m_{iz} &= \left(  \frac{k^2 c_{i}^2 - \omega^2}{c_{i}^2} \right)^{1/2}, \qquad \text{for } i=0,1,2. \label{hm0z}
\end{align}	
With the same notation, the expansion of the dispersion relation for body waves becomes
\begin{align}
H_1 &+ H_2 - H_{0b} + \frac{1}{\gamma \omega^2} \left\{  \left[ 2 k^2 c_1^2 - \omega^2  \right] \frac{H_1}{\beta_1} +  \left[ 2 k^2 c_2^2 - \omega^2  \right] \frac{H2}{\beta_2} \right.  \nonumber \\
& \quad \left. - \frac{H_{0b}}{\beta_0} \left[  2 k^2 c_0^2 - \omega^2 \right] - \frac{2 x_0 n_{0z}^2}{\beta_0}  \left[1 + \binom{\mathrm{tan}^2}{\mathrm{cot}^2}\{ n_{0z} x_0\} \right] \right\} =0,
\end{align}	
where further
\begin{align}
H_{0b} &= \frac{2 n_{0z}}{ \omega^2} \binom{-\tan{}}{\cot{}} \{n_{0z} x_0\}, \\
n_{0z} &= \left(  \frac{ \omega^2 - k^2 c_{i}^2}{c_{i}^2} \right)^{1/2}. \label{hn0z}
\end{align}	
\par Let us now solve the dispersion relation for a few interesting cases of high-$\beta$ slabs, enclosed in high-$\beta$ environments, and visualise the solutions. Panel (a) of Figure \ref{fig:hhh} illustrates the results of the numerical examination in a typical high-$\beta$ equilibrium configuration. There is a band of fast body modes (corresponding to case (\ref{fba})) confined between the sound speeds, and a band of slow body modes between the internal Alfv{\'e}n- and cusp speeds (which represents the conditions outlined in (\ref{sba}). Here, slow surface waves are present as well, as opposed to the low-$\beta$ limit.
\par Next, panel (b) of Figure \ref{fig:hhh} shows that the dispersion curves do not change qualitatively when the Alfv{\'e}n speeds are interchanged. Besides the slow surface mode, there is still a band of fast body modes fulfilling the conditions of (\ref{fba}), and a band of slow body modes representative of (\ref{sbc}). However, when the sound speeds are interchanged, as it may be seen in panel (c) of Figure \ref{fig:hhh}, only the slow surface waves and the band of slow body waves appear, while there are no fast waves present at all.
\par The splitting of body mode bands remains allowed in the high-$\beta$ limit. Slow body modes adhering to the conditions in (\ref{sbc}), as well as slow surface modes are present. One band of fast body modes is confined between the internal sound speed and the lowest of the external cusp speeds, as outlined in (\ref{fbc}). A second band of fast body modes realizes case (\ref{fbb}), comprising of waves with  $v_{A1} < v_{ph} < c_{T2}$, while a third band of fast body modes corresponds to case (\ref{fba}) and contains waves with phase speeds $v_{A2} < v_{ph} < c_1$.

\subsection{Infinite-$\beta$ limit}		\label{sec:Infinitebeta} 

In this limit, magnetic forces can be considered negligible as compared to kinetic ones, and so the approximation $ v_{Ai} \approx 0$ for $i= 0, 1, 2$ can be taken, and only "fast" (i.e. essentially purely acoustic) body waves occur. The modified wavenumber coefficients simplify to the expressions of Equations \eqref{hm0z} and \eqref{hn0z}, and the first-order terms vanish from the dispersion relation. Using the pressure balance condition \eqref{eq:ratio}, the dispersion relation for body modes reduces to
\begin{align}
\binom{\tan}{-\cot} \{n_0 x_0\} = \frac{1}{2} \left( \frac{m_1}{n_0}   \frac{ c_1^2}{c_0^2} +  \frac{m_2}{n_0}   \frac{ c_2^2}{c_0^2} \right). \label{infinitebeta_body}
\end{align}
In the symmetric case, Equation \eqref{infinitebeta_body} further simplifies to Equations (24) and (25) of \citeauthor{roberts3} (\citeyear{roberts3}). The condition $n_{0z}$, $m_{1z}$ and $m_{2z}$ $>$ $0$ is only fulfilled when $k^2 c_{0}^2 < \omega^2 < \min{(k^2 c_{1}^2, k^2 c_{2}^2)}$.  The band of fast body waves therefore exists between the internal sound speed and the lower of the two external ones. Similarly to the zero-$\beta$ case, an infinite number of harmonics exist in the direction of structuring due to the periodicity of the tangent and cotangent functions. Introducing the notation $ c_m=\min{(c_{1}, c_{2})}$, the waves are expected to behave as $\omega^2 = k^2 c_{m}^2  [\rho_m/\rho_0] \left[ 1 + \nu / (kx_0)^2 \right]$. By using the alternate method described during the derivation of the general wide-slab approximations, the coefficients $\nu_j$ can be determined. Eventually, the quasi-sausage mode solutions are given as
\begin{align}
\omega^2   = k^2 c_{m}^2 \frac{\rho_m}{\rho_0} \left[ 1 + \frac{\pi^2 \left( j - \frac{1}{2}\right)^2}{k^2 x_0^2} \right], \label{infinitebetas2}
\end{align}
while the approximation for quasi-kink modes becomes
\begin{align}
\omega^2  = k^2 c_{m}^2 \frac{\rho_m}{\rho_0} \left[ 1 + \frac{\pi^2 j^2}{k^2 x_0^2} \right]. \label{infinitebetak2}
\end{align}
In this case, Equations \eqref{infinitebetas2} and \eqref{infinitebetak2} showcase a simple connection between the lower external sound speed, and the ratio of the same side's external density to the internal one for any given value of the wavenumber and angular frequency of a given order body mode.
Similarly to the low-$\beta$ case, in the limits of high and infinite plasma-$\beta$ as well, the asymmetry brings a more complex dependence of the frequencies of eigenmodes on the set of external parameters characteristic of the system. The difference of external equilibrium parameters affects the frequencies of surface-, as well as body waves, and introduces cut-off frequencies regarding to the trapped propagation of both. Notably, due to this cut-off, there can be more than one band of either fast or slow body modes. Furthermore, in the wide-slab limit, the phase speeds of surface modes will diverge. The latter can lead to a phenomenon known as avoided crossing, which will be detailed in the next section.

\section{The effect of varying magnetic field and density ratios}		\label{sec:avcross}

   \begin{figure}  
\centering
\includegraphics[width=0.75\textwidth, height=0.35\textwidth, keepaspectratio]{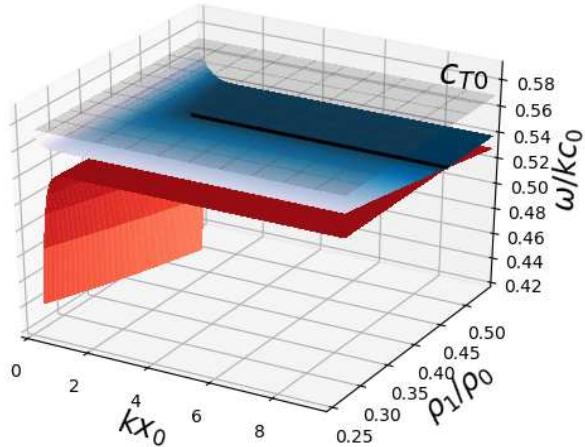}
\caption{The slow quasi-sausage and quasi-kink surface mode solutions of the dispersion relation are plotted for a fixed value of dimensionless slab width ($kx_0$), and changing density ratio on one side of the slab. The other density ratio using the density from the other side of the slab is held fixed at $\rho_2/\rho_0 = 0.4$. The characteristic speed orderings are identical to those of Figure \ref{fig:hhh}, but $c_1$ varies to satisfy equilibrium pressure balance. The black bold line indicates the values of the density ratio and the dimensionless slab width, for which the phase speeds of the quasi-sausage and quasi-kink modes perform a close approach and avoided crossing.}
   \label{fig:avcross}              
   \end{figure}

\begin{sloppypar} Avoided crossings of eigenmodes are known to happen in various physical processes, from quantum mechanics through coupled spring oscillations, to photochemistry \citep{ naqv, deva, heiss, nov}. In MHD, they were first found on dispersion diagrams of magneto-acoustic gravity waves of a plane stratified atmosphere by \citeauthor{abdel} (\citeyear{abdel}), and further examined by e.g. \citeauthor{mather} (\citeyear{mather}). Avoided crossings occur when constraints in a physical system supporting wave perturbations preclude the phase speeds of two modes from being equal, which is accompanied by a transferral of properties between the modes. \citeauthor{asymm} (\citeyear{asymm}) showed that avoided crossing happens between quasi-sausage and quasi-kink modes of a slab in a non-magnetic asymmetric environment, when the density ratio of the two external domains is varied.  \end{sloppypar}
\par In the current study, we find that the quasi-sausage and quasi-kink eigenmodes of an asymmetric slab in a magnetic environment perform avoiding crossings as well. Figure \ref{fig:avcross} demonstrates this phenomenon for the slow surface modes under the equilibrium conditions used in Figure \ref{fig:hhh}b. This behaviour is not specific to slow mode solutions, but since the fast surface mode does not exist in a high-$\beta$ configuration, our examination proceeds with the slow surface modes.
\par A substantial difference from the non-magnetic case is that the closest approach between the phase speeds of the slow quasi-sausage and quasi-kink surface modes does not occur at equal external densities this time, due to the presence of the magnetic asymmetry. Keeping the external Alfv{\'e}n speed $v_{A1}$ fixed while varying the external density ratio $\rho_1/\rho_0$ implies that the strength of the external equilibrium magnetic field $B_1$ is continuously changing throughout this numerical examination, too. Thus, the case of equal external densities ($\rho_1=\rho_2$) on its own does not correspond to a symmetric configuration, and the phase speeds of the quasi-modes will show the greatest similarity at a different value of the changing density ratio.
\par
\begin{figure}    
                               
   \centerline{\hspace*{0.001\textwidth}
               \includegraphics[width=0.9\textwidth,height=0.65\textwidth,clip=]{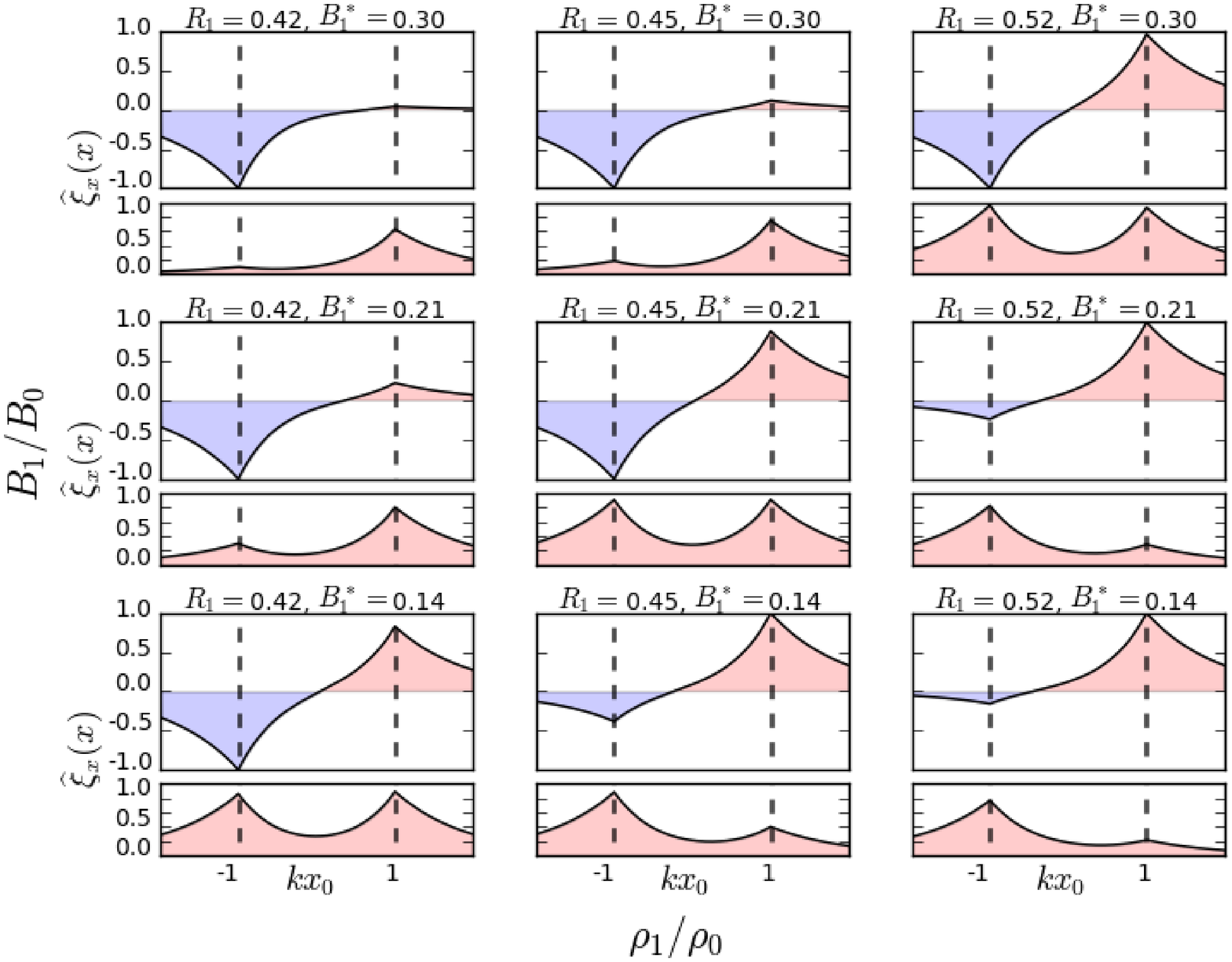}
              }
     \vspace{-0.6\textwidth}   
     \centerline{\Large \bf     
      \hspace{0.05 \textwidth}  \color{black}{(a)}
         \hfill}
     \vspace{0.58\textwidth}    
   \centerline{\hspace*{0.001\textwidth}
			\includegraphics[width=0.8\textwidth,height=0.55\textwidth,clip=]{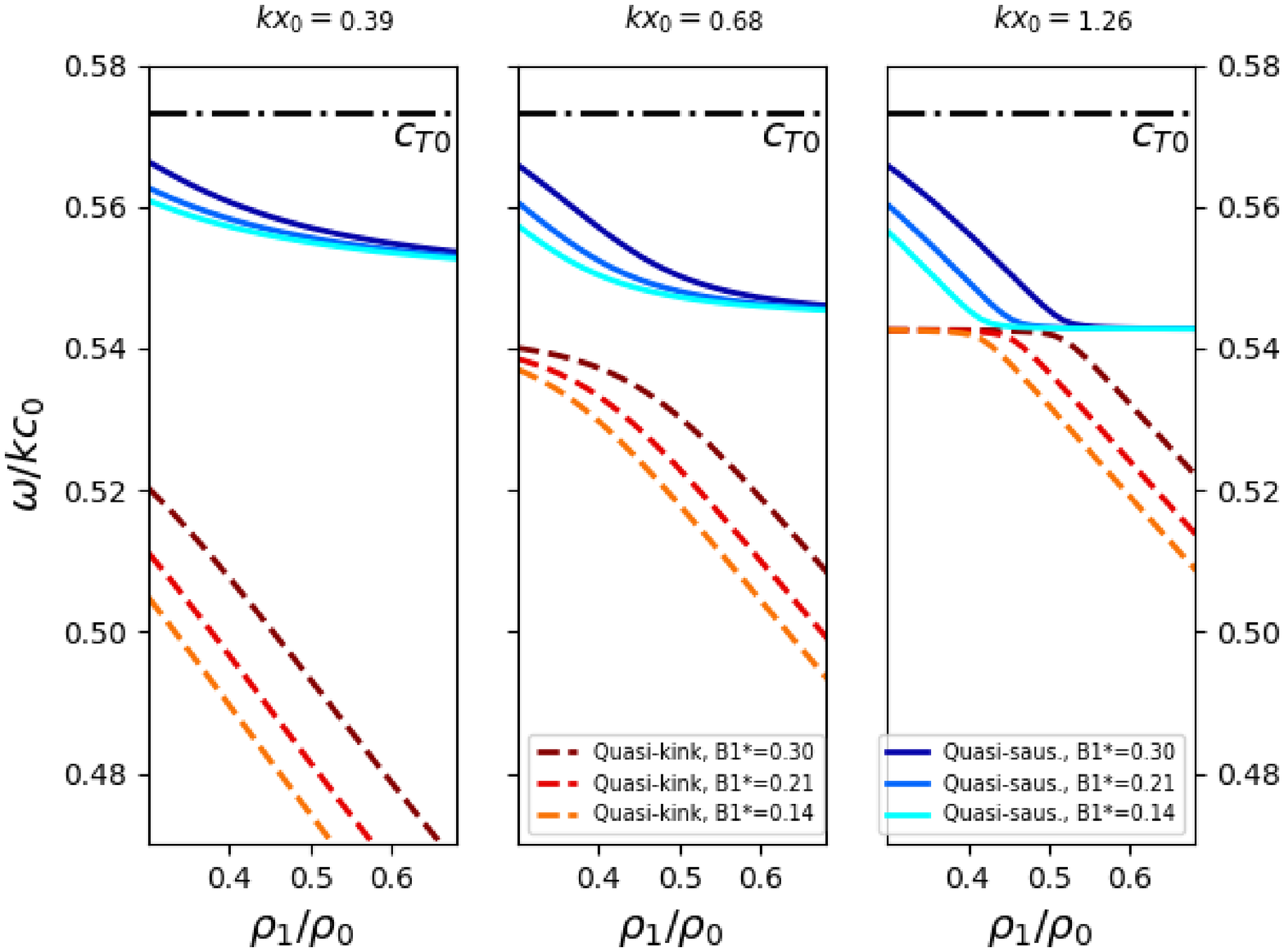}
              }
     \vspace{-0.52\textwidth}   
     \centerline{\Large \bf     
      \hspace{0.05 \textwidth}  \color{black}{(b)}
         \hfill}
     \vspace{0.48\textwidth}    
\caption{(a) The spatial variation of the transverse displacement perturbation ($\hat{\xi_x}$) is plotted. The upper (lower) parts of the panel represent the quasi-sausage (quasi-kink) mode solutions. In each column, the left-side density ratio remains constant, while in each row, the ratio of the left-side external magnetic field to the internal one ($B_1^{*} = B_1/B_0$) is kept at the same value. The right-side density ratio is held fixed at $\rho_2/\rho_0 = 0.4$. The characteristic speeds are:  $v_{A0}=0.7 c_0$, $v_{A1}=0.2 c_0$, $v_{A2}=0.1 c_0$, $c_{2}=1.8742 c_0$, but $c_1$ varies to satisfy equilibrium pressure balance. Panel (b) displays solution curves corresponding to different values of $B_1^{*}$, for specific values of the non-dimensional slab width ($kx_0$).}
   \label{fig:avcross3}              
   \end{figure}

 It may be seen in Figures \ref{fig:avcross}-\ref{fig:avcross3}, avoided crossings happen when either the density ratio on one side, or the ratio between one of the external equilibrium magnetic field stength values to the internal one is changed. In the figure presented, the left-side external Alfv{\'e}n speed, $v_{A1}$, grows from the lower right to the upper left corner. The displacement perturbations of quasi-sausage and quasi-kink modes, as a result, show the effect of avoided crossing, as one follows the diagonal from the first, through the fifth, to the ninth panel. Figure \ref{fig:avcross3}b illustrates how the changing magnetic field ratio shifts the point of closest approach for different $kx_0$ values.
\par We conclude that, although both thermodynamic and magnetic asymmetry can cause avoided crossings to occur, the behaviour of the slow quasi-sausage and quasi-kink modes during such approaches is qualitatively similar as in the case of an asymmetric slab with a field-free environment. The consecutive panels in the rows of Figure \ref{fig:avcross3}a show that as the symmetric configuration is approached, the amplitude of the quasi-sausage mode on the two interfaces begins to change, and the plane with the highest amplitude eventually shifts from the left side to the right, following the interface with the lower density ratio. In the meantime, the highest amplitude of the quasi-kink mode does the exact opposite, by jumping from the right to the left boundary of the slab, thus following the interface with the higher density ratio. The same exchange of properties can also be observed in the columns of Figure \ref{fig:avcross3}a, this time governed by the relative magnitudes of the external magnetic fields.

\section{Conclusion}		\label{sec:conc}

\par Wave dispersion in a magnetic slab embedded in plasma atmospheres of various structures (magnetic or free of field, uniform or asymmetric) is a complex problem that has been studied for decades, and yet still offers new solutions and discoveries. The associated dispersion relation for wave propagation, in general, is a transcendental equation. The dispersion relation often describes a rich spectrum of normal modes. Investigating a magnetic slab surrounded by an asymmetric field-free environment, \citeauthor{asymm} (\citeyear{asymm}) found that the difference in external conditions leads to important changes in wave dispersion. All of the solutions are described by one and the same dispersion relation, and the eigenmodes show mixed characteristics. 
\par The situation is qualitatively similar in the case of added magnetic asymmetry in the environment. After deriving the equation that governs wave dispersion in this configuration, and examining the incompressible limit \citep{asymag}, we have now continued to explore various approximations in important and limiting cases. With the aim of providing the theoretical background for future applications, analytically solvable equations descriptive of wave behaviour were retrieved for slabs much thinner or wider than the characteristic length-scale set by the wavelength of perturbations. The presence of a magnetically asymmetric environment modifies the frequencies of eigenmodes, and introduces a number of cut-off frequencies, as well as new possibilities for the ordering of characteristic speeds, and therefore, different phase speed bands in which trapped solutions remain possible. All these various new and interesting cases deserve their own description, since the analytical expressions retrieved in the thin- and wide-slab, as well as low- and high-$\beta$ limits simplify the calculations to be performed. Furthermore, they provide clear connections between the physical parameters describing the system, and the properties (wavenumbers, angular frequencies) of eigenmodes, which express the influence of environmental asymmetry. 
\par With these approximations, thus, a set of mathematical tools is provided, that we can use to describe a plethora of asymmetric solar astrophysical waveguides, such as e.g. the global stratification of the solar atmosphere, prominences or plumes in the corona, and magnetic bright points, light bridges or light walls in the photosphere. While these are all promising candidates to apply our asymmetric slab model to, we emphasize that there are natural limitations to the applicability. The validity of considering a solar structure as a slab sandwiched between asymmetric external layers is case-dependent and determined first and foremost by the extent of local gradients in plasma/magnetic parameters. Using an asymmetric slab model to describe a solar structure is a sensible approach if the difference between the three regions constructed is relatively big compared to the variation of background parameters within the three regions (which are essentially averaged out in this description). Therefore, the spatial scale of local gradients in the direction of structuring (i.e. the $x$-direction) should be comparable to the size of the slab. This assumption may or may not be true in general; it should be evaluated on a case-by-case basis for the specific waveguides one intends to study.
\par For a thin slab, most of the solutions are analogous to the supported modes of a slab placed in a symmetric magnetic environment. There are, however, a few more possibilities to arrange characteristic speeds, not all of which can be attributed to a direct parallel with a simple symmetrisation of external parameters. For a first approximation for body modes, the asymmetry mainly shows as quantitative modifications and cut-offs in their frequency, beyond which the modes would become leaky. The ratio of the internal density to the external ones directly appears in the description of surface waves, while it does not appear in the approximation for body modes. This is pointing to the fact that the latter are less sensitive to changes in the density ratio.
\par We have also examined how the ratio of plasma kinetic and magnetic pressures affects supported modes. These approximations can serve as the basis of direct applications to solar physics, which is to be the subject of a follow-up article. Here, it was detailed how, in a more general high-$\beta$ environment, representative of photospheric circumstances, all but the fast surface mode solutions might appear. However, under upper-chromospheric/coronal conditions, when the plasma-$\beta$ is low in all three domains, only body waves are present. 
\par The model becomes even more adaptable by combining the equations of geometrical and plasma-$\beta$ approximations, and provides analytical solutions for various structures in the solar atmosphere which can be handled as a slab. For example, the region of coronal hole boundaries might be thought of as an asymmetric magnetic slab, and plumes have already been reported to show MHD perturbations.

\acknowledgments
Acknowledgements: All numerical results are derived using Python, an open-source and community-developed programming language. The authors thank M. Allcock and M. Barbulescu for the basis of the root-finding algorithm used during the numerical investigation. The authors also acknowledge the support received from the Erasmus Programme of the EU for enabling the start of this research. N. Zs{\'a}mberger is also grateful to the University of Debrecen, E{\"o}tv{\"o}s Lor{\'a}nd University and the University of Sheffield. R. Erd{\'e}lyi is grateful to Science and Technology Facilities Council (STFC, grant numbers ST/M000826/1) for the support received. R.Erd{\'e}lyi also acknowledges the support received by the CAS Presidents International Fellowship Initiative Grant No. 2019VMA052 and the warm hospitality received at USTC of CAS, Hefei, where part of his contribution was made.

\appendix

\section{Quasi-sausage surface mode solutions in the thin-slab limit} 		\label{sec:appendix-tube}

The surface wave solution in the thin-slab approximation, with the approach $\omega^2 \Rightarrow k^2 c_{T0}^2$, can exist as trapped when the conditions in \eqref{ctconditions} are met. However, if this is not the case, the following possibilities exist, depending on the ordering of characteristic speeds in the different plasma layers:
\begin{align}
&\omega^2 = k^2 c_{T0}^2 \left[ 1 + \frac{2 (c_0^2 - c_{T0}^2) (v_{A1}^2-c_{T0}^2)^{1/2} (v_{A2}^2-c_{T0}^2)^{1/2} k x_0}{\rho_0 v_{A0}^2 c_0^2 R_v} \right], \nonumber \\
&R_v = \frac{1}{\rho_2} \frac{(v_{A1}^2-c_{T0}^2)^{1/2} (c_{T0}^2-c_{2}^2)^{1/2}}{(v_{A2}^2+c_{2}^2)^{1/2} (c_{T0}^2-c_{T2}^2)^{1/2}} + \frac{1}{\rho_1} \frac{(v_{A2}^2-c_{T0}^2)^{1/2} (c_{T0}^2-c_{1}^2)^{1/2}}{(v_{A1}^2+c_{1}^2)^{1/2} (c_{T0}^2-c_{T1}^2)^{1/2}},  \\
&\text{for} \qquad  c_1^2 < c_{T0}^2 <v_{A1}^2 \qquad \text{and } \qquad c_2^2 < c_{T0}^2 <v_{A2}^2; \nonumber
\end{align}
\begin{align}
&\omega^2 = k^2 c_{T0}^2 \left[ 1 - \frac{2 (c_0^2 - c_{T0}^2) (c_{T0}^2-v_{A1}^2)^{1/2} (c_{T0}^2-v_{A2}^2)^{1/2} k x_0}{\rho_0 v_{A0}^2 c_0^2 R_v } \right], \nonumber \\
&R_v= \frac{1}{\rho_2} \frac{(c_{T0}^2-v_{A1}^2)^{1/2} (c_{2}^2-c_{T0}^2)^{1/2}}{(v_{A2}^2+c_{2}^2)^{1/2} (c_{T0}^2-c_{T2}^2)^{1/2}} + \frac{1}{\rho_1} \frac{(c_{T0}^2-v_{A2}^2)^{1/2} (c_{1}^2-c_{T0}^2)^{1/2}}{(v_{A1}^2+c_{1}^2)^{1/2} (c_{T0}^2-c_{T1}^2)^{1/2}}  \label{sausage3}\\
& \text{for} \qquad  v_{A1}^2 < c_{T0}^2 <c_1^2 \qquad \text{and } \qquad v_{A2}^2 < c_{T0}^2 < c_2^2. \nonumber \end{align}
\par All three cases lead back to the results that were obtained for the symmetric slab in a magnetic environment, see e.g. \citeauthor{roberts3} (\citeyear{roberts3}), equations (16b, c, d) accordingly, if we substitute $v_{A1}^2=v_{A2}^2=v_{Ae}^2$, $c_1^2=c_2^2=c_e^2$, $\rho_1 = \rho_2 = \rho_e$.
\par With different orderings of characteristic speeds, the asymmetry may give rise to solutions other than those listed above. The following cases might also occur:
\begin{align}
&\omega^2 = k^2 c_{T0}^2 \left[ 1 + \frac{2 (c_0^2 - c_{T0}^2) (v_{A1}^2-c_{T0}^2)^{1/2} (v_{A2}^2-c_{T0}^2)^{1/2} k x_0}{\rho_0 v_{A0}^2 c_0^2 R_v} \right], \nonumber \\
&R_v = \frac{1}{\rho_2} \frac{(v_{A1}^2-c_{T0}^2)^{1/2} (c_{T0}^2-c_{2}^2)^{1/2}}{(v_{A2}^2+c_{2}^2)^{1/2} (c_{T0}^2-c_{T2}^2)^{1/2}} + \frac{1}{\rho_1} \frac{(v_{A2}^2-c_{T0}^2)^{1/2} (c_{1}^2-c_{T0}^2)^{1/2}}{(v_{A1}^2+c_{1}^2)^{1/2} (c_{T1}^2-c_{T0}^2)^{1/2}},  \\
&\text{for} \qquad  c_{T0}^2 <c_{T1}^2 \qquad \text{and } \qquad c_2^2 < c_{T0}^2 <v_{A2}^2; \nonumber \label{asymthinsurfsaus1}
\end{align}
\begin{align}
&\omega^2 = k^2 c_{T0}^2 \left[ 1 + \frac{2 (c_0^2 - c_{T0}^2) (v_{A1}^2-c_{T0}^2)^{1/2} (c_{T0}^2-v_{A2}^2)^{1/2} k x_0}{\rho_0 v_{A0}^2 c_0^2 R_v} \right], \nonumber \\
&R_v = - \frac{1}{\rho_2} \frac{(v_{A1}^2-c_{T0}^2)^{1/2} (c_{2}^2-c_{T0}^2)^{1/2}}{(v_{A2}^2+c_{2}^2)^{1/2} (c_{T0}^2-c_{T2}^2)^{1/2}} + \frac{1}{\rho_1} \frac{(c_{T0}^2-v_{A2}^2)^{1/2} (c_{1}^2-c_{T0}^2)^{1/2}}{(v_{A1}^2+c_{1}^2)^{1/2} (c_{T1}^2-c_{T0}^2)^{1/2}},  \\
&\text{for} \qquad  c_{T0}^2 <c_{T1}^2 \qquad \text{and } \qquad  v_{A2}^2< c_{T0}^2 <c_2^2; \nonumber \label{asymthinsurfsaus2}
\end{align}
\begin{align}
&\omega^2 = k^2 c_{T0}^2 \left[ 1 - \frac{2 (c_0^2 - c_{T0}^2) (v_{A1}^2-c_{T0}^2)^{1/2} (c_{T0}^2-v_{A2}^2)^{1/2} k x_0}{\rho_0 v_{A0}^2 c_0^2 R_v} \right], \nonumber \\
&R_v = \frac{1}{\rho_2} \frac{(v_{A1}^2-c_{T0}^2)^{1/2} (c_{2}^2-c_{T0}^2)^{1/2}}{(v_{A2}^2+c_{2}^2)^{1/2} (c_{T0}^2-c_{T2}^2)^{1/2}} - \frac{1}{\rho_1} \frac{(c_{T0}^2-v_{A2}^2)^{1/2} (c_{T0}^2-c_{1}^2)^{1/2}}{(v_{A1}^2+c_{1}^2)^{1/2} (c_{T0}^2-c_{T1}^2)^{1/2}},  \\
&\text{for} \qquad  c_1^2 < c_{T0}^2 <v_{A1}^2 \qquad \text{and } \qquad v_{A2}^2 < c_{T0}^2 <c_{2}^2.  \nonumber \label{asymthinsurfsaus3}
\end{align}
Solutions (\ref{asymthinsurfsaus1}), (\ref{asymthinsurfsaus2}), and (\ref{asymthinsurfsaus3}) also provide a mathematical description for cases where the same conditions are met, only the $i=1,2$ indices of all the characteristic speeds and densities are swapped. To tie them to a symmetric equivalent, however, a ''dominant'' condition needs to be chosen for these last three configurations (for example, by choosing the average of the two external parameters as the value in the symmetric case, and using this average to set the conditions). Such a treatment reveals that Equations (\ref{asymthinsurfsaus1}) and (\ref{asymthinsurfsaus2}) reduce to equation (16b) of \citeauthor{roberts3} (\citeyear{roberts3}) if the dominant condition is $c_{T0}<c_{Te}$. Equation (\ref{asymthinsurfsaus1}) reduces to \citeauthor{roberts3}'s equation (16c) with the condition choice $c_e^2 < c_{T0}^2 < v_{Ae}^2$, while relation (\ref{asymthinsurfsaus2}) reduces to their equation (16d) if the dominant condition is  $v_{Ae}^2 < c_{T0}^2 < c_e^2$. Equation (\ref{asymthinsurfsaus3}), however, does not have an equivalent in the symmetric case with either possible ordering of characteristic speeds for that case.

\section{Thin- and wide-slab approximations based on the full dispersion relation} 	\label{sec:appendix-width}

\par It is possible to utilise the equilibrium information that the slab is thin, and obtain the quasi-sausage mode solutions for the full dispersion relation (\ref{surface}) as well. Keeping in mind that in the thin-slab approximation, both $\tanh{(m_0 x_0)} \rightarrow m_0 x_0$ and $\coth{(m_0 x_0)} \rightarrow 1/(m_0 x_0)$ are valid, we can examine each mode mentioned in Section \ref{sec:ThinSurface}. 
 For the quasi-sausage mode whose phase speed approaches $c_{T0}$ in the limit of a thin slab, using the substitution $\omega^2 \rightarrow k^2 c_{T0}^2$, the full dispersion relation can be rearranged into a third-degree equation of the form
\begin{align}
	{U}^{{3}}+{a}_{{2}}{U}^{{2}}+{a}_{{0}}=0, 
\end{align}
where
\begin{align*}
	{a}_{{2}}&=2\frac {{k}^{{3}}{W}_{{1}}{W}_{{2}}V A_{{0}}}{E}, &{a}_{{0}}&=\frac {2{k}^{{5}}{v}^{{3}}{A}_{{0}}{A}_{{1}}^{{2}}{A}_{{2}}^{{2}}{R}_{{1}}{R}_{{2}}{x}_{{0}}}{E}+{V}^{{3}}{A}_{{0}}{k}^{{4}}{x}_{{0}}^{{2}},\\
	A_0&= \sqrt{(v_{A0}^2-c_{T0}^2)}, 	&A_1&= \sqrt{(v_{A1}^2-c_{T0}^2)},\\
	A_2&= \sqrt{(v_{A2}^2-c_{T0}^2)},  &E&= R_1 A_1^2 W_2 + R_2 A_2^2 W_1,\\
	 R_1&= \frac{\rho_1}{\rho_0}, &R_2&= \frac{\rho_2}{\rho_0}, \\
	V&= \sqrt{\frac{(c_{0}^2-c_{T0}^2)}{(v_{A0}^2+c_{0}^2)}}, 	&W_1&= \sqrt{\frac{(v_{A1}^2-c_{T0}^2)(c_{1}^2-c_{T0}^2)}{(v_{A1}^2+c_{1}^2)(c_{T1}^2-c_{T0}^2)}},\\
	 W_2&= \sqrt{\frac{(v_{A2}^2-c_{T0}^2)(c_{2}^2-c_{T0}^2)}{(v_{A2}^2+c_{2}^2)(c_{T2}^2-c_{T0}^2)}}, &U&=\sqrt{{{k}^{{2}}{c}_{T0}^{{2}}-{\omega }^{{2}}}}.
\end{align*}
The (real) solutions are then given by
\begin{align}
	{\omega }^{{2}}={k}^{{2}}{c}_{{T0}}^{{2}}- \left( \frac{S}{6} + \frac{2 a_2^2}{3 S} -\frac{a_2}{3} \right)^2,
\end{align}
where
\begin{align*}
	S= \left( -108 a_0 - 8 a_2^2 + 12 \sqrt{12 a_0 a_2^3 + 81 a_0^2} \right)^{1/3}.
\end{align*}
\par In order to keep the solutions real, the same six orderings of the characteristic speeds are possible that were detailed using the decoupled dispersion relation.
\par Through a similar process, the quasi-sausage mode with $\omega^2 \rightarrow k^2 c_2^2$ can be obtained from the full dispersion relation as:
\begin{align}
\omega^2 = k^2 c_2^2 - \left( \frac{2k^2x_0 R_1 R_2 A_1 A_2 V_0^2 + k^3 x_0^2 R_2 A_0^2 A_2 V_0^2 V_1 + k R_2 A_2 V_1}{2k x_0 A_0^2 V_1 W + k^2 x_0^2 R_1 A_0^2 A_1 V_0^2 W + R_1 A_1 W}  \right)^2,
\end{align}
where
\begin{align*}
	A_0&= \sqrt{(v_{A0}^2-c_{2}^2)}, &A_1&= \sqrt{(v_{A1}^2-c_{2}^2)},\\
	A_2&= \sqrt{(v_{A2}^2-c_{2}^2)}, &W&= \sqrt{\frac{1}{(c_{T2}^2 - c_2^2)(c_2^2 + v_{A2}^2)}}, \\
	R_1&= \frac{\rho_1}{\rho_0}, &R_2&= \frac{\rho_2}{\rho_0}, \\
	V_0&= \sqrt{\frac{(c_0^2-c_2^2)}{(c_{T0}^2-c_{2}^2) (v_{A0}^2 + c_0^2)}}, &V_1&= \sqrt{\frac{(c_1^2-c_2^2)}{(c_{T1}^2-c_{2}^2) (v_{A1}^2 + c_1^2)}}.
\end{align*}
For a more symmetric configuration, the dispersion relation for the quasi-sausage mode approaching $c_1=c_2=c_e$ simplifies to the following equation:
\begin{align}
	2{A}_{{0}}^{{2}}{W}_{{1}}{W}_2 x_0 {U}^{{2}} +E\left({1+{A}_{{0}}^{{2}}{V}^{{2}}{k}^{{2}}{x}_{{0}}^{{2}}}\right)U+2{R}_{{1}}{R}_{{2}}{v}^{{2}}{A}_{{1}}{A}_{{2}}{x}_{{0}}=0,
\end{align}
where
\begin{align*}
	A_0&= \sqrt{(v_{A0}^2-c_{e}^2)}, &A_1&= \sqrt{(v_{A1}^2-c_{e}^2)},\\
	A_2&= \sqrt{(v_{A2}^2-c_{e}^2)}, &E&= R_1 A_1 W_2 + R_2 A_2 W_1, \\
	R_1&= \frac{\rho_1}{\rho_0}, &R_2&= \frac{\rho_2}{\rho_0}, \\
	U&=\sqrt{{{k}^{{2}}{c}_{e}^{{2}}-{\omega }^{{2}}}}, &V&= \sqrt{\frac{(c_{0}^2-c_{e}^2)}{(v_{A0}^2+c_{0}^2)(c_{T0}^2-c_{e}^2)}}, \\
	W_1&= \sqrt{\frac{1}{(v_{A1}^2+c_{1}^2)(c_{T1}^2-c_{e}^2)}}, &W_2&= \sqrt{\frac{1}{(v_{A2}^2+c_{2}^2)(c_{T2}^2-c_{e}^2)}}.
\end{align*}

\par For the quasi-kink mode with $\omega^2 \rightarrow k^2 v_{A1}^2$, the solutions become
\begin{align}
\omega^2 = k^2 v_{A1}^2 - \left(  \frac{2 k^2 x_0 A_0^2 A_2 W_1 W_2 + k^3 x_0^2 R_2 A_0^2 A_2^2 V^2 W_1 + k R_2 A_2^2 W_1}{2 k x_0 R_1 R_2 A_2^2 V^2 + k^2 x_0^2 R_1 A_0^2 A_2 V^2 W_2 + R_1 A_2 W_2)}  \right)^2,
\end{align}
where
\begin{align*}
	A_0&= \sqrt{(v_{A0}^2-v_{A1}^2)}, &A_2&= \sqrt{(v_{A2}^2-v_{A1}^2)},\\
	R_1&= \frac{\rho_1}{\rho_0}, &R_2&= \frac{\rho_2}{\rho_0}, \\
	W_1&= \sqrt{\frac{(c_1^2 - v_{A1}^2)}{(c_{T1}^2 -v_{A1}^2)(c_1^2 + v_{A1}^2)}}, &W_2&= \sqrt{\frac{(c_2^2-v_{A1}^2)}{(c_{T2}^2-v_{A1}^2) (v_{A2}^2 + c_2^2)}}, \\
	V&= \sqrt{\frac{(c_0^2-v_{A1}^2)}{(c_{T0}^2-v_{A1}^2) (v_{A0}^2 + c_0^2)}}.
\end{align*}
For the even more symmetric case, when the phase speed is approaching $v_{A1}=v_{A2} = v_{Ae}$, the dispersion relation for the quasi-kink mode simplifies to
\begin{align}
	2{R}_{{1}}{R}_{{2}}{V}^{{2}}{x}_{{0}}^2 U^2+E\left[{{k}^{{2}}{V}^{{2}}{A}_{{0}}^{{2}}{x}_{{0}}^{{2}}+1}\right]U+2{W}_{{1}}{W}_{{2}}{A}_{{0}}^{{2}}{k}^{{2}}{x}_{{0}}=0,
\end{align}
where
\begin{align*}
	A_0&= \sqrt{(v_{A0}^2-v_{Ae}^2)}, 	&E&= R_1 W_2 + R_2  W_1, \\
	R_1&= \frac{\rho_1}{\rho_0}, &R_2&= \frac{\rho_2}{\rho_0}, \\
	U&=\sqrt{{{k}^{{2}}{v}_{Ae}^{{2}}-{\omega }^{{2}}}}, &V&= \sqrt{\frac{(c_{0}^2-v_{Ae}^2)}{(v_{A0}^2+c_{0}^2)(c_{T0}^2-v_{Ae}^2)}}, \\
	W_1&= \sqrt{\frac{(c_{1}^2-v_{Ae}^2)}{(v_{Ae}^2+c_{1}^2)(c_{T1}^2-v_{Ae}^2)}}, &	W_2&= \sqrt{\frac{(c_{2}^2-v_{Ae}^2)}{(v_{Ae}^2+c_{2}^2)(c_{T2}^2-v_{Ae}^2)}}.
\end{align*}
For another type of asymmetric quasi-kink mode, namely, with $\omega^2 \rightarrow k^2 c_{T1}^2$, the solutions become
\begin{align}
\omega^2 = k^2 c_{T1}^2 - \left(  \frac{2 k^2 x_0 A_0^2 W V_2 + k^3 x_0^2 R_2 A_0^2  A_2 V_0^2 W + k R_2 A_2 W}{2 k x_0 R_1 R_2 A_1 A_2 V_0^2 + k^2 x_0^2 R_1 A_0^2 A_1 V_0^2 V_2 +  R_1 A_1 V_2)}  \right)^2,
\end{align}
where
\begin{align*}
	A_0&= \sqrt{(v_{A0}^2-c_{T1}^2)}, &A_1&= \sqrt{(v_{A1}^2-c_{T1}^2)},\\
	A_2&= \sqrt{(v_{A2}^2-c_{T1}^2)}, &W&=  \sqrt{ \frac{(c_1^2 - c_{T1}^2)}{(c_1^2 + v_{A1}^2)}}\\
	R_1&= \frac{\rho_1}{\rho_0}, &R_2&= \frac{\rho_2}{\rho_0}, \\
	V_0&= \sqrt{\frac{(c_0^2 - c_{T1}^2)}{(v_{A0}^2+c_{0}^2)(c_{T0}^2-c_{T1}^2)}}, &V_2&= \sqrt{\frac{(c_2^2 - c_{T1}^2)}{(v_{A2}^2+c_{2}^2)(c_{T2}^2-c_{T1}^2)}}.
\end{align*}

\par Essentially, the same considerations apply for the derivation of body mode solutions starting from the full dispersion relation \eqref{fullbody}, as we used for the algebraic manipulations of the decoupled dispersion relation. For slow body waves in a thin slab, the angular frequency tends toward $|k c_{T0}|$ or the appropriate cut-off frequency (see the details after Equation \ref{sbgen}) from above, and $|n_0| \rightarrow \infty$. Now, for the expression $n_0 (-\tan{(n_0x_0)} + \cot{(n_0x_0)})$ to remain bounded as $kx_0 \rightarrow 0$, $n_0 x_0$ should approach values that satisfy $(-\tan{(n_0x_0)} + \cot{(n_0x_0)}) = 0$. The roots of this equation are the multiples of $\pi/4$, therefore 
\begin{align}
n_0 x_0  = \frac{2j-1}{4} \pi, \label{quarterpi}
\end{align}
where $j=1,2,3...$. Using this condition, the coefficients $\nu = \nu_j$ in Equation \eqref{sbgen} can be determined, and an approximation for $\omega^2$ provided, which will describe both quasi-sausage and quasi-kink modes. In a wide slab, slow body modes are still described by the approximation of Equation \eqref{stypeawide}. The angular freqency will therefore approach $|k\min{(v_{A0}, c_0)}|$ (potentially with some offset) from below. Taking this into consideration, in a high-beta slab, the same condition as in \eqref{quarterpi} can be set. However, if the slab is filled with low-beta plasma, then the coefficients $\nu_j$ for quasi-sausage and quasi-kink modes can be determined by ensuring that $(-\tan{(n_0x_0)} + \cot{(n_0x_0)}) \rightarrow \pm \infty$, which means for integers $j$ that
\begin{align}
n_0 x_0  = \frac{j}{2} \pi. \label{halfpi}
\end{align}
\par Similarly, fast body waves in a thin slab can be described by Equation \eqref{fbgen}. To determine the coefficients $\nu_j$, the condition \eqref{quarterpi} applies if the slab is in a low-beta environment, and Equation \eqref{halfpi} should be used in the case of a high-beta environment. In the wide slab, the approximation in Equation \eqref{fastwide} can still describe fast body waves, and the values of $\nu_j$ can be calculated by prescribing the condition \eqref{quarterpi} if $v_{A0} > c_0$, and fulfilling Equation \eqref{halfpi} if the opposite is true.

\section{Low- and high-$\beta$ approximations based on the full dispersion relation} 	\label{sec:appendix-beta}

Utilising the modified wavenumber coefficients (\ref{lowcoeff1}) and (\ref{lowcoeff2}), the full dispersion relation for surface waves can be expanded (to first order) to the following form for a configuration in which the plasma-$\beta$ is low everywhere:
\begin{align}
&2 L_{0A} + 2 m_{0z}^2 L_{0B} + m_{0z} L_{0C} \left\{ \tau_{0z} + \frac{1}{\tau_{0z}} \right\}   - \frac{1}{2} \bigg\{  m_{0z} L_{0C} \left[  \tau_{0z} + \frac{1}{\tau_{0z}} \right]  \left[ k^2 v_{A0}^2 - \omega^2 \right] - \frac{1}{2} \left[ \tau_{0z}^2 + \frac{1}{\tau_{0z}^2} \right] \left[ L_{0B} \right.   \nonumber \\
& \quad \left. +  x_0 \left( k^2 v_{A0}^2 - \omega^2 \right) L_{0C} \right] m_{0z}^2  + \left[ L_{0B} + \left( k^2 v_{A0}^2 - \omega^2 \right)  x_0 L_{0C} \right] m_{0z}^2  \bigg\} \gamma \beta_0   - \frac{1}{2} L_{0A} \gamma \left\{ \beta_1 + \beta_2 \right\} \nonumber \\
& \quad  - \frac{1}{2}  \gamma \rho_0 m_{0z}   \left\{ k^2 v_{A0}^2 - \omega^2 \right\} \left\{ \beta_1 \frac{m_{1z}}{\rho_1} \left[ k^2 v_{A2}^2 - \omega^2 \right] + \beta_2 \frac{m_{2z}}{\rho_2} \left[ k^2 v_{A1}^2 - \omega^2 \right] \right\} \left\{ \tau_{0z} + \frac{1}{\tau_{0z}} \right\}  = 0.
\end{align}
Here,
\begin{align}
L_{0A} &= \frac{\rho_0^2}{\rho_1 \rho_2} m_{1z} m_{2z} \left( k^2 v_{A0}^2 - \omega^2 \right),  \qquad &L_{0B} &= m_{0z}^2 \left( k^2 v_{A1}^2 - \omega^2 \right) \left( k^2 v_{A2}^2 - \omega^2 \right), \\
L_{0C} &= \rho_0 m_{0z} \left[ \frac{m_{1z}}{\rho_1} \left( k^2 v_{A2}^2 - \omega^2 \right) + \frac{m_{2z}}{\rho_2} \left( k^2 v_{A1}^2 - \omega^2 \right) \right], \qquad &\tau_{0z} &= \tanh{(m_{0z} x_0)}, \label{fullL}
\end{align}
and $m_{iz}$ ($i = 0,1,2$) are defined in Equation \eqref{lm0z}. For body waves, using the expressions from Equations \eqref{ln0z} and \eqref{fullL} and with $T_{0z}=\tan{(n_{0z}x_0)}$, the expansion, to first order, becomes:
\begin{align}
&2 L_{0A} - 2 n_{0z}^2 L_{0B} - n_{0z} L_{0C} \left\{ \frac{1}{T_{0z}} -  T_{0z}  \right\} + \frac{1}{2} \left\{  \frac{1}{2} \left[ \omega^2 \left( \frac{m_{1z}}{\rho_1} + \frac{m_2z}{\rho_2}\right) -1 \right]  \left[ \frac{1}{T_{0z}} - T_{0z} \right] \left[k^2 v_{A0}^2 - \omega^2 \right] \rho_0 n_{0z}  \right.  \nonumber \\
&\quad   \left.  +  \frac{1}{2} n_{0z} L_{0C} \left[  T_{0z}^2 + \frac{1}{T_{0z}^2} \right]  \left[ k^2 v_{A0}^2 - \omega^2 \right] + n_{0z}^2 \left[ x_0 \left( k^2 v_{A0}^2 - \omega^2 \right) L_{0c} + 2 L_{0B} \right] \right\} \gamma \beta_0 - \frac{1}{2} \gamma L_{0A} \left\{ k^2 v_{A0}^2 - \omega^2 \right\} \left\{ \beta_1 + \beta_2 \right\} \nonumber \\
& \quad - \frac{1}{4} \gamma \rho_0 n_{0z} \left\{ \frac{1}{T_{0z}} - T_{0z} \right\} \left\{ k^2 v_{A0}^2 - \omega^2 \right\} \left\{ \beta_1  \frac{m_{1z}}{\rho_1} \left[ k^2 v_{A2}^2 - \omega^2 \right] + \beta_2  \frac{m_{2z}}{\rho_2} \left[ k^2 v_{A1}^2 - \omega^2 \right] \right\} = 0.
\end{align}

Similarly, if the plasma-$\beta$ is high in all three domains, the expansion of the dispersion relation (to first order) is:
\begin{align}
&\left\{ 2 H_{0A} + 2 m_{0z}^2 + m_{0z} H_{0B} \left[ \tau_{0z} + \frac{1}{\tau_{0z}} \right] \right\} \omega^4 - \frac{2}{\gamma} k^2 \omega^2 \left\{ \left[  m_{0z} H_{0B} \left( \tau_{0z} + \frac{1}{\tau_{0z}}\right) \right. \right. \nonumber \\
& \quad \left. \left. + 4  H_{0A} \right] \frac{c_0^2}{\beta_0} + \rho_0 m_{0z} \left[  \frac{c_2^2}{\beta_2} \frac{m_{1z}}{\rho_1} + \frac{c_1^2}{\beta_1} \frac{m_{2z}}{\rho_2} \right] + 2 m_{0z}^2 \left[ \frac{c_1^2}{\beta_1} + \frac{c_2^2}{\beta_2} \right] \right\} = 0,
\end{align}
where 
\begin{align}
H_{0A} = \frac{\rho_0}{\rho_1} \frac{\rho_0}{\rho_2} m_{1z} m_{2z}, \qquad  \qquad H_{0B} = \rho_0 \left( \frac{m_{1z}}{\rho_1} + \frac{m_{2z}}{\rho_2}   \right), \label{fullH}
\end{align}
and $m_{iz}$ ($i=0,1,2$) are defined in Equation \eqref{hm0z}, and further $\tau_{0z}$ is defined in Equation \eqref{fullL}. Using the same factors, as well as Equation \eqref{hn0z}, and the notation $T_{0z}=\tan{(n_{0z}x_0)}$, the expanded full dispersion relation for body waves in a high-$\beta$ configuration can be written as
\begin{align}
&\left( 2 H_{0A} - 2 n_{0z}^2 - n_{0z} H_{0B} \left\{ T_{0z} - \frac{1}{T_{0z}}\right\}  \right) \omega^4 + \frac{\omega^2}{\gamma} \left( \bigg\{ \left[ n_{0z} H_{0B} \left( T_{0z}^2 + \frac{1}{T_{0z}^2} +2 \right)+ 2 k^2 c_0^2 \left( T_{0z} - \frac{1}{T_{0z}} \right) \right] \right.  \nonumber \\
& \quad  \left.   - 8 H_{0A} k^2 c_0^2 + 4 \omega^2 n_{0z}^2  \bigg\} \frac{1}{\beta_0} +2 k^2 \rho_0 n_{0z}\left\{ \frac{c_2^2}{\beta_2} \frac{m_{1z}}{\rho_1}  +\frac{c_1^2}{\beta_1} \frac{m_{2z}}{\rho_2}  \right\}  \left\{  T_{0z} - \frac{1}{ T_{0z}}\right\}  + 4 k^2 n_{0z}^2 \left\{ \frac{c_1^2}{\beta_1}  + \frac{c_2^2}{\beta_2}  \right\} \right) =0.
\end{align}

\bibliographystyle{aasjournal}
\bibliography{sola_magneticslab_bibl_paper2}


\end{document}